\RequirePackage{latexrelease}
\documentclass{aa}

\usepackage[varg]{txfonts}
\usepackage{graphicx}
\usepackage{array, booktabs, makecell}
\usepackage{siunitx}
\usepackage[version=4]{mhchem}
\usepackage{textcomp, gensymb}
\usepackage[export]{adjustbox}
\usepackage{amsmath}
\usepackage{amsfonts}
\usepackage{amssymb}
\usepackage{nicefrac}
\usepackage{comment}
\usepackage{multicol}
\usepackage[utf8]{inputenc}
\usepackage[english]{babel}
\usepackage{comment}
\usepackage[dvipsnames]{xcolor}
\makeatletter
\renewcommand*\aa@pageof{, page \thepage{} of \pageref*{LastPage}}
\makeatother

\usepackage{xcolor}
\usepackage{natbib,twoopt}
\usepackage[hyphenbreaks]{breakurl}
\usepackage[breaklinks]{hyperref}      %% to avoid \citeads line fills, add "draft" 
\bibpunct{(}{)}{;}{a}{}{,}             %% natbib format for A&A and ApJ
\definecolor{cobalt}{rgb}{0.06, 0.2, 0.65}
\hypersetup{
  colorlinks,
  citecolor=blue,
  linkcolor=[rgb]{0.8, 0.2, 1.0},
  urlcolor=blue,
}
\makeatletter
  \newcommandtwoopt{\citeads}[3][][]{\href{http://adsabs.harvard.edu/abs/#3}%
    {\def\hyper@linkstart##1##2{}%
     \let\hyper@linkend\@empty\citealp[#1][#2]{#3}}}
  \newcommandtwoopt{\citepads}[3][][]{\href{http://adsabs.harvard.edu/abs/#3}%
    {\def\hyper@linkstart##1##2{}%
     \let\hyper@linkend\@empty\citep[#1][#2]{#3}}}
  \newcommandtwoopt{\citetads}[3][][]{\href{http://adsabs.harvard.edu/abs/#3}%
    {\def\hyper@linkstart##1##2{}%
     \let\hyper@linkend\@empty\citet[#1][#2]{#3}}}
  \newcommandtwoopt{\citeyearads}[3][][]%
    {\href{http://adsabs.harvard.edu/abs/#3}
    {\def\hyper@linkstart##1##2{}%
     \let\hyper@linkend\@empty\citeyear[#1][#2]{#3}}}
\makeatother

\usepackage{orcidlink}

\begin{document}

\title{Probing anisotropic particle acceleration and limb-brightening in Centaurus A's jet} 
   
\author{Felix Glaser\inst{\ref{inst1}}\orcidlink{0009-0005-5870-4195}, Christian.~M.~Fromm\inst{\ref{inst1},}\inst{\ref{inst2}}\orcidlink{0000-0002-1827-1656}, Luca Ricci\inst{\ref{inst1}}\orcidlink{0000-0002-4175-3194}, Yosuke Mizuno\inst{\ref{inst4},}\inst{\ref{inst5},}\inst{\ref{inst6}}\orcidlink{0000-0002-8131-6730}, Matthias Kadler\inst{\ref{inst1}}\orcidlink{0000-0001-5606-6154},\\ Karl Mannheim\inst{\ref{inst1}}\orcidlink{0000-0002-2950-6641}, and Michael Janssen\inst{\ref{inst7},\ref{inst3}}\orcidlink{0000-0001-8685-6544}}
\institute{
Institute for Theoretical Physics and Astrophysics, University of Würzburg, Emil-Hilb-Weg 31, 97074 Würzburg, Germany \label{inst1}
\and Institute for Theoretical Physics, Goethe Universit\"at Frankfurt, Max-von-Laue-Str. 1, 60438 Frankfurt, Germany \label{inst2}
\and Tsung-Dao Lee Institute, Shanghai Jiao Tong University, Shanghai, 201210, People’s Republic of China \label{inst4}
\and School of Physics and Astronomy, Shanghai Jiao Tong University, Shanghai, 200240, People’s Republic of China \label{inst5}
\and Key Laboratory for Particle Astrophysics and Cosmology (MOE) and Shanghai Key Laboratory for Particle Physics and Cosmology, Shanghai Jiao Tong University, Shanghai 200240, People’s Republic of China \label{inst6}
\and Department of Astrophysics, Institute for Mathematics, Astrophysics and Particle Physics (IMAPP), Radboud University, P.O. Box 9010, 6500 GL Nijmegen, The Netherlands \label{inst7} 
\and Max-Planck-Institut für Radioastronomie, Auf dem Hügel 69, 53121 Bonn, Germany \label{inst3}\\
\email{felix.glaser@stud-mail.uni-wuerzburg.de,christian.fromm@uni-wuerzburg.de}}

\date{Received / Accepted}

\abstract
% context heading (optional)
{Relativistic jets are among the most fascinating objects in the Universe, and recent high-resolution Very Long Baseline Interferometric (VLBI) observations, including the Global mm-VLBI Array and the Event Horizon Telescope (EHT), are able to resolve their structure close to their launching site. These observations reveal strongly limb-brightened jet structures for Centaurus A (Cen\,A), M\,87 and 3C\,84. }
  % aims heading (mandatory)
{Thus, the question arises which physical mechanism can generate the limb-brightened structure, and if this structure is common for jets from low-luminosity active galactic nuclei (LLAGN) seen under large viewing angles. Therefore, as a pilot study, we aim to model the EHT observations of Cen\,A.}
  % methods heading (mandatory)
{We performed a 3D two-temperature general-relativistic magnetohydrodynamic (GRMHD) simulation of an accreting supermassive black hole (SMBH) and jet launching to study the plasma dynamics and computed the connected emission via general relativistic radiative transfer (GRRT) calculations considering possible anisotropies in the distribution of the radiating particles. In order to adjust our simulations to the EHT observations of Cen\,A, we carried out a Bayesian fitting in the Fourier plane.}
  % results heading (mandatory)
{We find that GRMHD simulations of magnetically arrested disks (MADs) combined with anisotropically emitting particle distributions along the direction of the magnetic field, parametrized by a value $\eta=0.07$, are able to mimic the recent EHT observations of Cen\,A. In addition, we extracted a black hole mass of $M_{\rm BH}=6\times10^7\,M_\odot$ and a viewing angle of $\vartheta=72^\circ$. Our obtained model can reproduce key features of the EHT and Atacama Large Millimeter/submillimeter Array (ALMA) observations in total and polarized emission. Finally, we predict that the black hole shadow in Cen\,A will be observable at a frequency of $\sim$ 3\,THz and stress the importance of high-cadence and high-frequency ($\nu>350$\,GHz) observations to better constrain the particle acceleration, radiation microphysics, and accretion dynamics.}
% conclusions heading (optional)
 {}  
\titlerunning{Probing anisotropic particle acceleration and limb-brightening in Centaurus A's jet}
\authorrunning{Glaser F., Fromm C. M. et al.}
\maketitle

\section{Introduction}
Black holes (BHs) are among the most fascinating objects in the universe, and the most massive ones with masses of millions to billions of suns are located in the center of galaxies. These so-called supermassive black holes (SMBHs) accrete matter from its surrounding, leading to the formation of an accretion disk. During the accretion process, roughly half of the matter's potential energy is released in the form of radiation across the entire electromagnetic spectrum, and these compact regions are referred as active galactic nuclei (AGN). Interestingly, around 10\% of the AGNs launch powerful and highly collimated relativistic outflows perpendicular to their disk, so-called jets. The physical processes connected to the accretion of matter onto SMBHs, the detailed emission processes as well as the formation of relativistic jets \citep{BZ1977,BP1982}, remain enigmatic despite decades-long observational and theoretical efforts.
\noindent
The closest AGN to Earth is located in the constellation Centaurus at a distance of $D=(3.8\pm0.1)\,{\rm Mpc}$ \citep{Harris_2010} and harbors a supermassive black hole with $M_{\rm BH}=(5.5\pm3.0)\times10^7\,M_\odot$ \citep{Neumayer_2010} which translates to a projected angular scale of $\sim0.14/\sin{(\vartheta)}$ gravitational radii $\left( r_g=GM_{\rm BH}/c^2\simeq8.1\times10^{12}\,\rm{cm}; \,\vartheta\equiv \rm{viewing\,angle}\right)$ per ${\rm \mu as}$. The highest angular resolution of $\sim 25\,{\rm \mu as}$ is provided by the Event Horizon Telescope at an observing wavelength of $\lambda=1.3$\,mm, and the observation of Centaurus\,A (Cen\,A) revealed a strongly edge-brightened jet \citep{Janssen2021}. Interestingly, the reconstructed image of Cen\,A showed both, a bright forward jet (pointing towards positive RA direction) and a dimmer counter-jet (pointing towards negative RA direction), where in both structures no emission was detected within the central spine. The detailed analysis of the jet width, $W$, revealed a "near-parabolic" shape with distance from the apex, $z$, i.e. $W\propto z^{0.33}$ while the jet-to-counter-jet ratio from the total surface brightnesses of ${R_{\rm j/cj}}\geq 5$ and from the brightest gaussian components ${R=1.6\pm0.5}$ suggest an viewing angle $\vartheta\neq90^\circ$. Combining the jet-to-counter-jet ratio with jet velocities obtained at larger wavelength led to estimates for the viewing angle range between $12^\circ\leq \vartheta \leq 45^\circ$ \citep{Mueller2014} in centimeter-wavelengths. On larger, but still sub-parsec scales, the jet is also found to exhibit a range of possible viewing angles between $50^\circ \leq \theta \leq  80^\circ$ \cite[][]{Tingay_1998}. Overall, the jet inclination is uncertain and varies strongly on different length scales and observing wavelengths of the jet, suggesting a more complex underlying jet structure. The values for the viewing angle naturally explain the jet-to-counter-jet ratio via Doppler boosting in the forward jet and de-boosting for the counter-jet \citep{Janssen2021}. Clearly, these observed features challenge both our theoretical understanding of jet physics and its numerical modeling. In this work, we aim to shed light on the detected limb-brightening in Cen\,A by combining state-of-the-art general-relativistic magnetohydrodynamics (GRMHD) simulations with advanced general-relativistic radiative transfer (GRRT) calculations, including pitch angle-dependent anisotropic emission connected to particle acceleration by magnetic reconnection.
\noindent
This paper is organized as follows: In Sect.~\ref{sec:theory} we briefly reconcile the observed image features with our theoretical expectations. Our numerical setup is presented in Sect.~\ref{sec: Numerical Modelling}, and the Bayesian fitting method is described in Sect.~\ref{sec:MCMC}. We provide the results of our modeling in Sect.~\ref{sec:results} followed by the discussion and summary in Sect.~\ref{sec:discussion} and Sect.~\ref{sec:summary}.

%\begin{comment}
\section{Theoretical considerations}
\label{sec:theory}
According to our understanding, accreting supermassive black holes are surrounded by bright rings of light caused by the strong gravitational lensing effects. The angular radius of these "photon orbits", assuming a non-spinning black hole, $a_\star$ is given by:
\begin{equation}
\Theta_{\rm photon}=\frac{\sqrt{27}GM_{\rm BH}}{c^2D}\simeq0.74\left(\frac{M_{\rm BH}}{5.5\times10^7\,M_\odot}\right)\left(\frac{D}{3.8\,{\rm Mpc}}\right)\,{\rm \mu as} 
\label{eq:photonorbit}
\end{equation}
Photons propagating at smaller distances than $\Theta_{\rm photon}$ will be captured by the black hole. The combined effects of the hot plasma and, therefore, glowing accretion disk, the emission from the photons on the photon orbits, and the photons lost to the black hole inside it, lead to the formation of the so-called "black hole shadow". Clearly, an angular resolution of $\lesssim 1\,{\rm \mu as}$ is required for the detection and imaging of the photon orbit in Cen\,A, which is not achievable with current radio interferometers. However, future space-based VLBI arrays at higher frequencies ($\nu>230\,\mathrm{GHz}$) than the EHT maybe be able to provide the required angular resolution and therefore resolve the black hole shadow in Cen\,A. Therefore, we do not expect to image and resolve the black hole shadow in the EHT observations of Cen\,A in contrast to the EHT observations of M87 \citep{EHT_M87_P1}.\\
\noindent
Regarding the structure and the nature of the jets in general, and in Centaurus A in particular, there are three launching scenarios: i) jet launching from the ergosphere of a rotating black hole typically referred to the Blandford-Znajek (BZ) mechanism \citep{BZ1977}, ii) the launching from the rotating accretion disk labeled as the Blandford-Payne (BP) mechanism \citep{BP1982} or iii) a combination of both processes. In addition to the detailed launching process, the structure and physics of the accretion disk plays a major role in the formation of the jets. The mass accretion rate determines the structure and physics of the accretion disk. Given a mass accretion rate of $\sim10^{-4}\dot{M}_{\rm Edd}$ \citep{Rothschild2011} Cen\,A can be classified as a low luminosity AGN and the accretion flow is assumed to be advection dominated accretion flows (ADAF) leading to geometrically thick and optically thin disks \citep{Meier2012}.   
Distinguishing these three scenarios would require detecting and resolving the jet footpoints, which are most likely located on scales of $\sim 10\,r_{\rm g}$\footnote{Assuming the anchoring of the jet in ergosphere for BZ and on ISCO scales for BP mechanism}. However, the EHT provides an angular resolution of $25\,{\rm \mu as}$ at an observing frequency of 230\,GHz to physical scales of $\sim 200\,r_{\rm g}$ for Centaurus A. Thus, the possible jet footpoints in Centaurus A cannot be resolved.\\ \\
\noindent
Nevertheless, the observed jet structure can be used to infer the physical conditions in the jet. Following the hypothesis that the jet in Cen\,A is launched by the BZ mechanisms, we expect, based on numerical simulations, the generation of a stratified jet with a dilute, fast, and highly magnetized spine and a dense, slower, and low magnetized sheath \citep[see, e.g.,][]{Fromm2022}. Within such a spine-sheath jet, the observed edge-brightening could be explained by the following mechanisms: 
\begin{itemize}
    \item Due to relativistic aberration, the emission of highly relativistic electrons will be boosted into a narrow cone with an opening angle of $\sim 1/\Gamma$ where $\Gamma$ is the bulk Lorentz factor of the electrons. Therefore, if the viewing angle is larger than the boosting cone, the emission will be de-boosted \citep[see, e.g.,][]{Pacholczyk1970}. %Given the estimates for the viewing angle $12\leq \vartheta \leq 45$ the required Lorentz factor for the particles in the spine to be de-boosted is 4.7 ($\vartheta=12^\circ$) and 1.2 ($\vartheta=45^\circ$).
    \item Given the large magnetic fields in the spine, the emitting electrons will cool down faster than in the sheath ($d\gamma_e/d\tau\propto -\gamma_e^2B^2$). Thus, the jet will appear empty as compared to a less cooled sheath. %If we assume that most of the radiation is emitted at the critical frequency $\nu_{\rm crit,obs}$, the electron Lorentz factor can be written as $\gamma_{\rm e,crit}=\sqrt{2\pi m_e c\nu_{\rm crit,obs}/(eB\delta)}$ where $m_e$ is the electron mass, $e$ the electron charge and $\delta$ the Doppler factor. Once the electron Lorentz factor is above $\gamma_{\rm e,crit}$ the jet spine will appear "empty", due to the radiative cooling. 
    \item Particles accelerated by magnetic reconnection develop a pitch angle anisotropy, which suppresses the commonly assumed isotropic distribution of pitch angle \cite[e.g.][]{Comisso2024}. This pitch angle anisotropy generates anisotropic electron distribution functions \citep{Galishnikova2023} leading to strongly edge-brightened jets \citep{Tsunetoe_2025}.
\end{itemize}

\section{Numerical modeling}\label{sec: Numerical Modelling}
\subsection{General-relativistic magnetohydrodynamics}
\label{sec:GRMHD}
We employ the 3D GRMHD code \textit{Kokkos-based High-Accuracy Relativistic Magnetohydrodynamics with AMR} (\texttt{KHARMA}; \cite{Prather2021, prather2024}), which simulates accretion onto and jet launching from black holes.
\texttt{KHARMA} solves the ideal, non-radiative MHD equations on curved space-times with four-metric $g_{\mu\nu}$ and metric determinant $g$ in geometric units ($G=c=1$ and a factor of $\nicefrac{1}{\sqrt{4\pi}}$ is absorbed in the magnetic field). Throughout this paper we will use the code units for the time- and length-scales, i.e. the light-crossing time is redefined to $t_\mathrm{g} = 1\, G M_\mathrm{BH}/(c^3) \equiv 1\, \unit{M}$ and the gravitational radius is $r_\mathrm{g} = 1\, G M_\mathrm{BH}/(c^2) \equiv 1\, \unit{M}$, where $G$ is Newton's constant, $c$ is the vacuum speed of light, $M_\mathrm{BH}$ is the BH mass and $\unit{M}$ is the mass unit that scales the units in the simulations. Greek indices run through $(0,1,2,3)$ and Roman indices through $(1,2,3)$, and we make use of the Einstein summation convention. We use a Kerr black hole solution on a modified spherical coordinate system (Funky-Modified-Kerr-Schild-coordinates; FMKS). These coordinates exhibit a decreasing grid cell size in the $\theta$-direction towards the equator ($\theta=\nicefrac{\pi}{2}$), which become increasingly cylindrical along the jet-axis ($\theta=0$ and $\theta=\pi$). The radial extent of the cells increases exponentially; high resolution is given at small radii, where finely resolving the accretion process is vital. ($r$, $\theta$, $\phi$) refer to ordinary spherical coordinates. We employed a simulation grid with $N_r =384 $, $N_\theta=192$, and $N_\phi=128$ grid cells in the respective directions. The radial boundaries range from $r = (1.1659-10000)\,\unit{M}$, and otherwise use the complete map in $\phi=(0-2\pi)$ and $\theta=(0-\pi)$. Our simulation is initialized with an FM-torus \citep{FM1976} in hydrostatic equilibrium characterized by its inner edge, $r_\mathrm{in}=20\,M$, the location of the maximum pressure, $r_\mathrm{c}=41\,M$, and its angular momentum, $l=6.88$ surrounding a black hole with spin $a_\star=0.94$. 
%The values for our torus are summarized in Table~\ref{tab:GRMHDinitial}. 
The initial torus is seeded with a weak poloidal magnetic field given by the vector potential
\begin{equation}\label{eq:vectorpotential}
	A_\phi = \max\left(\frac{\rho}{\rho_\mathrm{max}}\left(\frac{r}{r_\mathrm{in}}\right)^3\sin^3\theta\exp\left(\frac{-r}{400}\right)-0.01,0\right)
\end{equation}
and normalized to a plasma $\beta$ of $100$ \cite[see][for details on the impact of the vector-potential on the jet]{Glaser2026}. We assert an adiabatic index of $\hat{\gamma}=4/3$ to the fluid, assuming an ideal gas as equation of state.

\begin{figure*}[h!]
    \centering
    \includegraphics[width=0.99\linewidth]{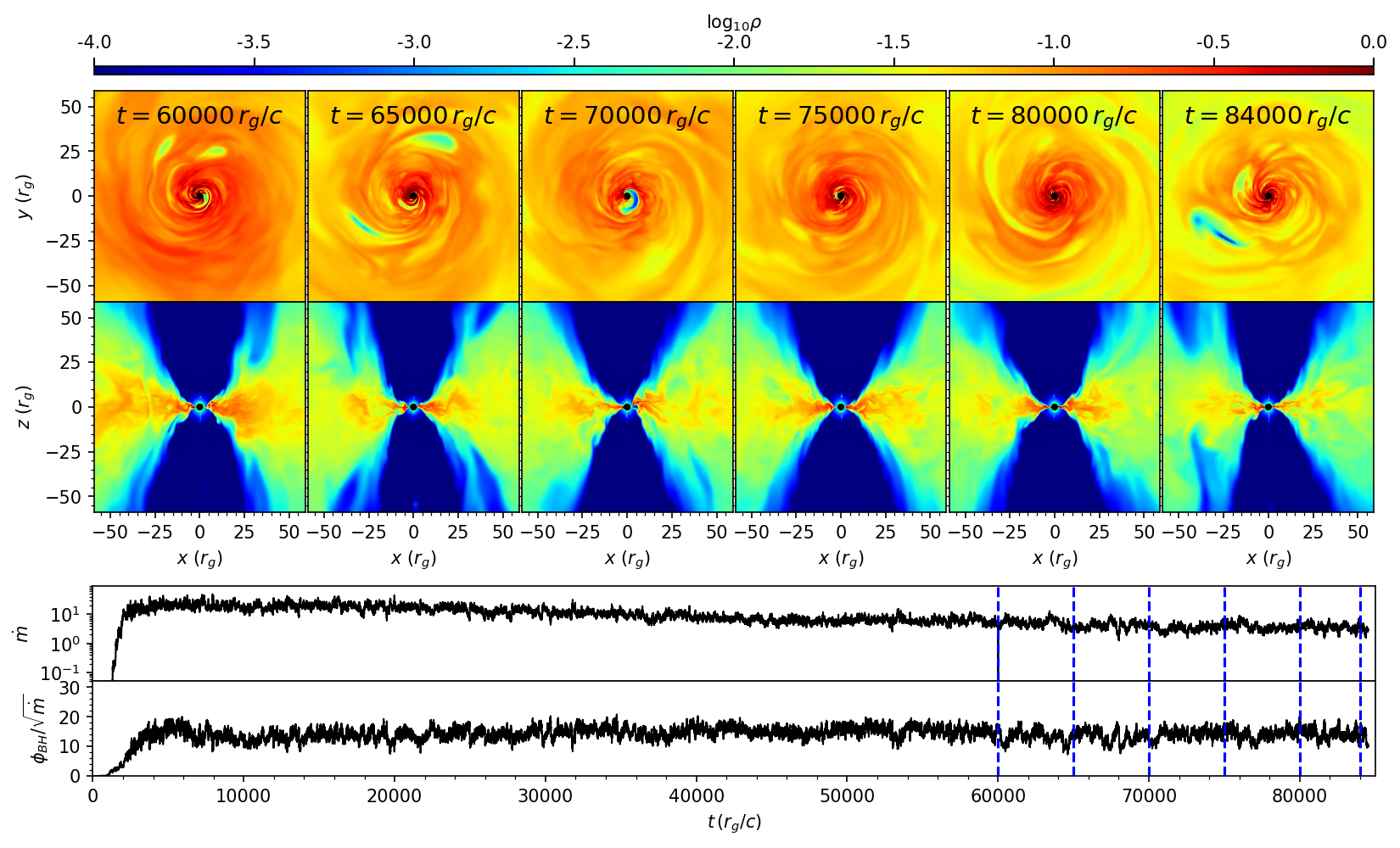}
    \caption{Evolution of the GRMHD simulation. The top and middle panels show the logarithm of the density in the equatorial and the meridional plane, respectively, for six different times focused on the late evolution. The bottom panel shows the temporal evolution of the mass accretion rate $\dot{m}$ and the MAD parameter $\Phi=\phi_{\rm BH}/\sqrt{\dot{m}}$. The blue vertical dashed lines mark the time frames of the density plots.}
    \label{fig:GRMHD}
\end{figure*}
\noindent
During the simulations, we evolve the electron entropy, $k_e$, separately using the turbulent heating fraction provided by \cite{Kawazura_2020}, which is based on compressively driven gyrokinetic turbulence \citep[see, e.g.][for details]{Mizuno2021}. To start the accretion onto the black hole, we perturb the internal energy by $u_g = u_g\left(1 + X_p\right)$ with a random number $\left|X_p\right | < 0.1$ to trigger the magneto-rotational instability (MRI). As usual in GRMHD simulations, we apply floors to the low-density fluid to ensure the stability of the numerical methods. In particular, we apply geometrical floors on the density $\rho_{\rm fl}=10^{-6}r^{-3/2}$ and on the internal energy $u_{\rm fl}=10^{-8}r^{-3/2\hat{\gamma}}$. We evolve our simulation for $t=85\,\rm kM$ and monitor the mass accretion rate, $\dot{m}$, and the magnetic flux across the horizon $\phi_{BH}$, which are defined following \citet{Porth2019} as:
\begin{equation}
    \dot{m}=\int^{2\pi}_0\int^{\pi}_0 \rho u^{r}\sqrt{-g}d\theta d\phi \quad \phi_{\rm BH}=\frac{1}{2}\int^{2\pi}_0\int^{\pi}_0 \left|B^r\right| \sqrt{-g}d\theta d\phi
\end{equation}

\noindent
In Fig.~\ref{fig:GRMHD} we present the temporal evolution of our GRMHD simulation. The top panel displays the evolution of the density in the equatorial plane, and the middle panel shows that in the meridional plane. The equatorial plane is dominated by high-density spiral accretion streams and low-density blobs ejected during MAD states (best seen at $t=70\,\mathrm{kM}$). In the meridional plane, the low-density funnel corresponds to the jet, and truncation of the accretion flow towards the black hole during the ejection of the MAD blobs is clearly visible as low-density gaps close to the black hole and vertical low-density strips further away from the black hole (see $t=70\,\mathrm{kM}$ and $t=84\,\mathrm{kM}$). The bottom panel shows the evolution of the mass accretion rate, $\dot{m}$, and the MAD parameter $\Phi=\phi_{\rm BH}/\sqrt{\dot{m}}$. The average MAD parameter is $\left<\Phi\right>= 14.4\pm2.6$, indicating that the MAD state is obtained throughout the course of the simulation \cite[][]{tchekhovskoy2011}.

\subsection{General-relativistic radiative transfer}\label{sec:GRRT}

In order to investigate the radiative signatures of the 3D GRMHD simulation, we calculate the polarized synchrotron emission by means of a general relativistic radiative transfer (GRRT) post-processing step. We therefore employ \texttt{IPOLE} \citep{Monika2018}: a semi-analytic code for relativistic polarized radiative transfer.

The electron temperature is computed from the electron entropy, $k_{\rm{e}}$ according to:
\begin{equation}
    \Theta_{\rm e} =k_{\rm e}\rho^{\hat{\gamma_{\rm e}}-1}\frac{m_{\rm p}}{m_{\rm e}},
\end{equation}
where $\hat{\gamma_{\rm e}}=4/3$ is the adiabatic index of the electrons and $m_{\rm p}$ is the proton mass. 
%\end{comment}
Along with this, the electron distribution function (eDF) is given by either a thermal (mainly in the disk) relativistic Maxwell-J\"{u}ttner distribution, whereas we model the non-thermal emission in the jet by a $\kappa$-eDF. The relativistic Maxwellian distribution is given by
\begin{equation}\label{eq: thermal-eDF}
     \frac{1}{n_e}\frac{dn_e}{d\gamma_\mathrm{e}}=\frac{\gamma_\mathrm{e}^2\sqrt{1-1/\gamma_\mathrm{e}^2}}{\Theta_e K_2(1/ \Theta_\mathrm{e})}\\
\end{equation}
where $\gamma_{\rm e}$ is the electron Lorentz factor, $\Theta_{\rm e}=k_{\rm B}T_{\rm e}/m_e c^2$ is the electron temperature, and $K_2$ is the modified Bessel function of the second kind. The non-thermal $\kappa$-eDF \citep[see][]{Pandya_2016}, which smoothly converges to the Maxwell-J\"{u}ttner distribution for low $\gamma_\mathrm{e}$ is given by:
 \begin{equation}\label{eq: kappa-eDF}
     \frac{1}{n_e}\frac{dn_e}{d\gamma_\mathrm{e}}=N\gamma_\mathrm{e}\sqrt{\gamma_\mathrm{e}^2-1}\left(1+\frac{\gamma_\mathrm{e}-1}{\kappa w}\right)^{-(\kappa+1)},
\end{equation}
with normalization factor $N$, power-law index $p=\kappa-1$ and width parameter $w$. 
We model the parameters $\kappa$ and $w$ with Particle-in-Cell (PIC)-based subgrid models. For the $\kappa$-value we use a prescription of \cite{Meringolo_2023}, who fit the parameter-space of the plasma-$\beta$ and magnetization $\sigma$, based on relativistic turbulent PIC simulations:  
\begin{equation}
    \kappa:=2.8+0.2\frac{1}{\sqrt{\sigma}}+1.6\sigma^{-6/10}\tanh{\left(2.25\sigma^{1/3}\beta \right)}
    \label{eq:kappavalue}
\end{equation}
The width parameter regulates the non-thermal excess compared to the Maxwellian, which is constituted by the magnetic energy-density available to the electrons \citep[][]{Davelaar2019,Fromm2022}:
\begin{equation}\label{eq: kappa width}
    w := \frac{\kappa-3}{\kappa}\Theta_\mathrm{e}+\frac{\varepsilon}{2}\left[1+\tanh{(r-r_\mathrm{inj})}\right]\frac{\kappa-3}{6\kappa}\frac{m_\mathrm{p}}{m_\mathrm{e}}\sigma,
\end{equation}
where $r_{\mathrm{inj}}$ is the injection radius of non-thermal particles, which is set to 10\,M, approximately the average location of the stagnation surface, where the injection could take place. 
This excess energy is regulated with a parametrization of $\varepsilon$ with $\sigma$ and plasma-$\beta$ also taken from \cite{Meringolo_2023}:
\begin{equation}
    \varepsilon = 1.0 - 0.23\frac{1}{\sqrt{\sigma}}+0.5\sigma^{1/10}\tanh{(-10.18\beta\sigma^{1/10})}
    \label{eq:kappaeff}
\end{equation}
The $\kappa$-eDF is only applied where $3.1<\kappa<8$, where the values calculated below the lower threshold are set to $\kappa=3.1$, and if the values exceed $\kappa=8$, the eDF switches to the thermal distribution, as the $\kappa$-eDF converges to the Maxwell-J\"{u}ttner eDF for high $\kappa$-values. In the first panel of Fig.~\ref{fig:emisspar}, we present the distribution of the $\kappa$ parameter, and in the second panel the distribution of $w$. Notice that $w$ involves the electron temperature obtained from turbulent heating and the efficiency parameter $\epsilon$.

\subsubsection{Anisotropic synchrotron emission}

It is well motivated, observationally and theoretically, that electrons or positrons in a weakly collisional or collisionless plasma in the presence of a dynamically strong magnetic field develop an anisotropic eDF w.r.t. the pitch angle. It is suggested that the pitch angle anisotropy is active in astrophysical scenarios such as the jet launching from BHs \citep{Zhdankin_2023}. One possible mechanism for the development of the anisotropic electron eDF is the so-called synchrotron firehose instability (SFHI) \citep{Zhdankin_2023}. Synchrotron cooling of relativistic electrons in an optically thin medium, even starting from an ab initio thermal equilibrium state, radiates away their kinetic energy perpendicular to the magnetic field lines, hence developing an anisotropic energy distribution. If the plasma-$\beta$ is sufficiently high, the plasma becomes unstable to the firehose instability. This then scatters the particles from low to high pitch angles.\\
\noindent
We model the anisotropic emission similarly to \cite{Tsunetoe_2025}, who introduced the case of the anisotropic eDF for a power-law distribution of relativistic electrons. In Sect.~\ref{sec:GRRT} we have introduced the isotropic $\kappa$-eDF. We can exploit the fact that the $\kappa$-eDF converges to a power-law eDF ${dn_e}/d\gamma_\mathrm{e}\propto \gamma_\mathrm{e}^{-p}$ for high Lorentz factors with a power-law index $p=\kappa-1$, consistent with the highly relativistic limit for the anisotropic power-law distribution in \cite{Tsunetoe_2025}, as well. The electron Lorentz factors in the jet are sufficiently high to support this approximation. The anisotropic emission of the highly relativistic electrons will be further beamed along the magnetic field lines; therefore, the electron pitch angle anisotropy can be expanded around the pitch angle $\theta_B$ between the wave vector of the radiation and the magnetic field vector. Hence we introduce the pitch angle anisotropy by expanding the isotropic, full-Stokes emissivities $j_{a,\,\rm{iso}}$ and absorption coefficients $\alpha_{a,\,\rm{iso}}$ with $a\in\{ I,Q,U \}$ around the pitch angle $\theta_B$ to yield the anisotropic pitch angle dependent emissivity and absorption coefficients \citep{Melrose_1971,Tsunetoe_2025}:

\begin{align}\label{eqn:anisoja}
    j_a&= j_{a,\,\rm{iso}}\, \phi(\theta_B), \quad
    &j_V= j_{V,\,\rm{iso}}\, \phi(\theta_B) \, \left( 1+ \frac{g(\theta_B)}{p+2}\right),\\
% \end{align}
% \begin{align}
    \alpha_a&= \alpha_{a,\,\rm{iso}}\, \phi(\theta_B),\quad
    &\alpha_V= \alpha_{V,\,\rm{iso}}\, \phi(\theta_B) \, \left( 1+ \frac{g(\theta_B)}{p+2}\right),
\end{align}
with 
\begin{align}
    \phi(\theta_B) &= P(p,\eta)^{-1} \left( 1+(\eta-1) cos^2(\theta_B) \right)^{-p/2},\\
    g(\theta_B) &= \frac{p(\eta-1)\sin^2(\theta_B)}{1+(\eta-1)\cos^2(\theta_B)}
\end{align}
and the normalization factor;
\begin{equation}
    P(p,\eta)=\int_0^1 \rm{d}\mu (1+(\eta-1)\mu^2)^{-p/2}.
    \label{eqn:peta}
\end{equation}
As mentioned above, we set $p$ consistently with the definition of the $\kappa$-parameter, i.e.,\ $p = \kappa - 1$ (see Fig.~\ref{fig:aniso}, where we illustrate the effect of pitch-angle  $\theta_{\rm B}$ on the degree of anisotropy of the emissivity/absorptivity distributions).

\subsection{Synchrotron cooling}
Electrons emitting synchrotron radiation are subject to radiative cooling. We mimic the radiative cooling of the non-thermal electrons by equating the synchrotron cooling time to the advection time scale, following \cite{Chatterjee2021}. Notice that this implementation of radiative cooling occurs in the post-processing GRRT calculations and does not affect the plasma dynamics. The break Lorentz factor, $\gamma_{\rm e,br}$ and corresponding break frequency, $\nu_{\rm br}$, i.e., the electron Lorentz factor, respectively, the frequency above which radiative cooling becomes important can be written as:
\begin{equation}
    \gamma_{\rm e,br}=\frac{6\pi m_{\rm e} c \left | v^r\right |}{\sigma_{\rm T} b^2 r}\quad \mathrm{and}\quad \nu_{\rm br}=\frac{3}{2}\frac{\gamma_{\rm e,br}^2 e b}{2\pi m_{\rm e} c},
    \label{eq:gbreak}
\end{equation}
where $v^r$ is the bulk fluid radial velocity, $\sigma_{\rm T}$ is the Thompson cross section and the magnetic field $b$. 

\begin{figure}[h!]
    \centering
    \includegraphics[width=\linewidth]{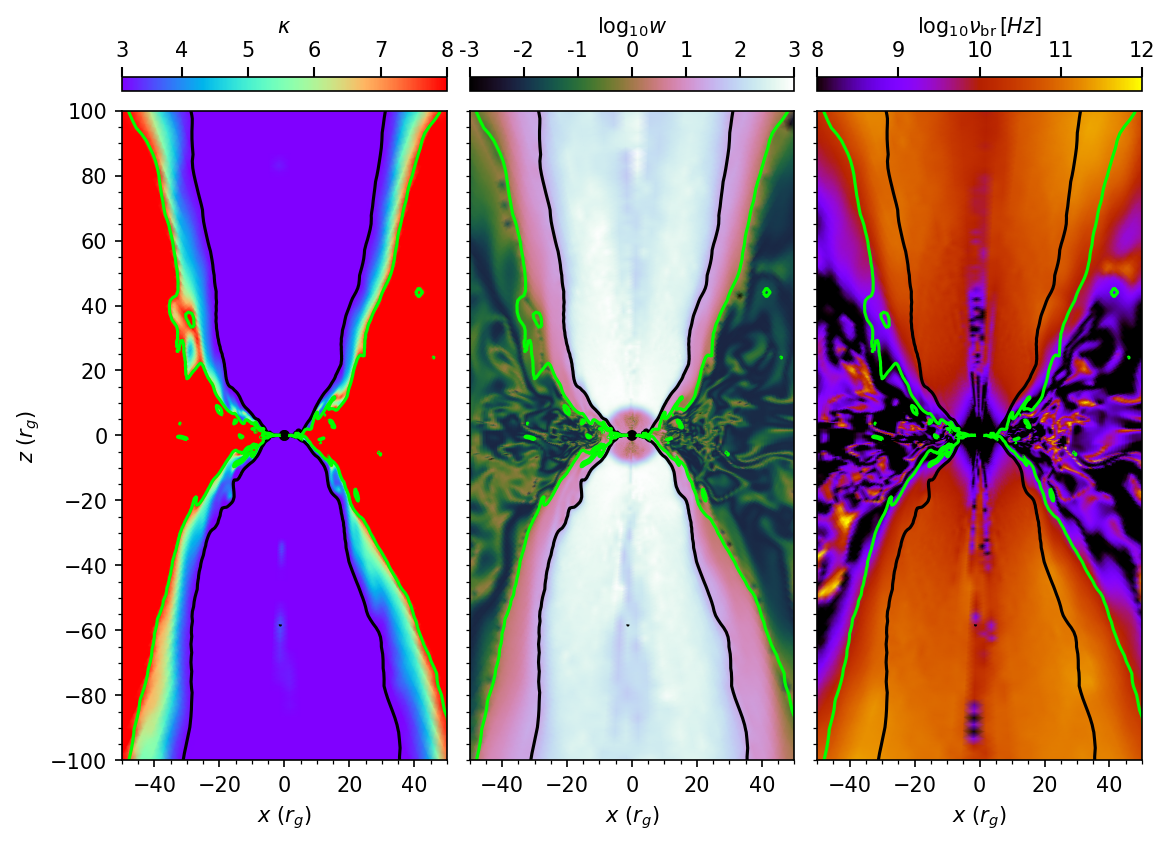}
    \caption{Distribution of the kappa parameter, $\kappa$, the width of the kappa-eDF, $w$, and the break frequency, $\nu_{\rm br}$, for $t=55\,\rm kM$. The black contour line corresponds to $\sigma=1$ and the green one to $\kappa=8$. Between these contours we apply the kappa-eDF during the radiative transfer calculations.} %\cf{provide the values for $M_{\rm BH}$ and $m$ simulations are not scale-free}}
    \label{fig:emisspar}
\end{figure}

In our calculations, we increase the parameter $\kappa$ by one $\left(\kappa \rightarrow \kappa+1\right)$ for frequencies above the break frequency and divide the emissivity by $\sqrt{\nu/\nu_{\rm break}}$ as in \cite{Scepi2022}. In Fig.~\ref{fig:emisspar} we show in the third panel the distribution of the break frequency $\nu_{\rm br}$ in the meridional plane for $t=55\,\mathrm{kM}$. Notice that radiative cooling is mainly important for regions within $r<40\,\rm M$. It is important to mention that by including radiative cooling even in the presented (over-) simplified form (as it does not feed back on the plasma dynamics in the GRMHD), the radiative transport calculations are no longer scale-free, i.e., the calculations need to be re-done if the black hole mass and/or the mass accretion rate (henceforth: mass unit; see Sect.~\ref{sec:MCMC}) is changed. This will become important during the synthetic data generation and Bayesian fitting (see Sect.~\ref{sec:MCMC}).

\section{Constraints from the EHT observations and Bayesian fitting}
\label{sec:MCMC}

\begin{figure}[h!]
    \centering
    \includegraphics[width=\linewidth]{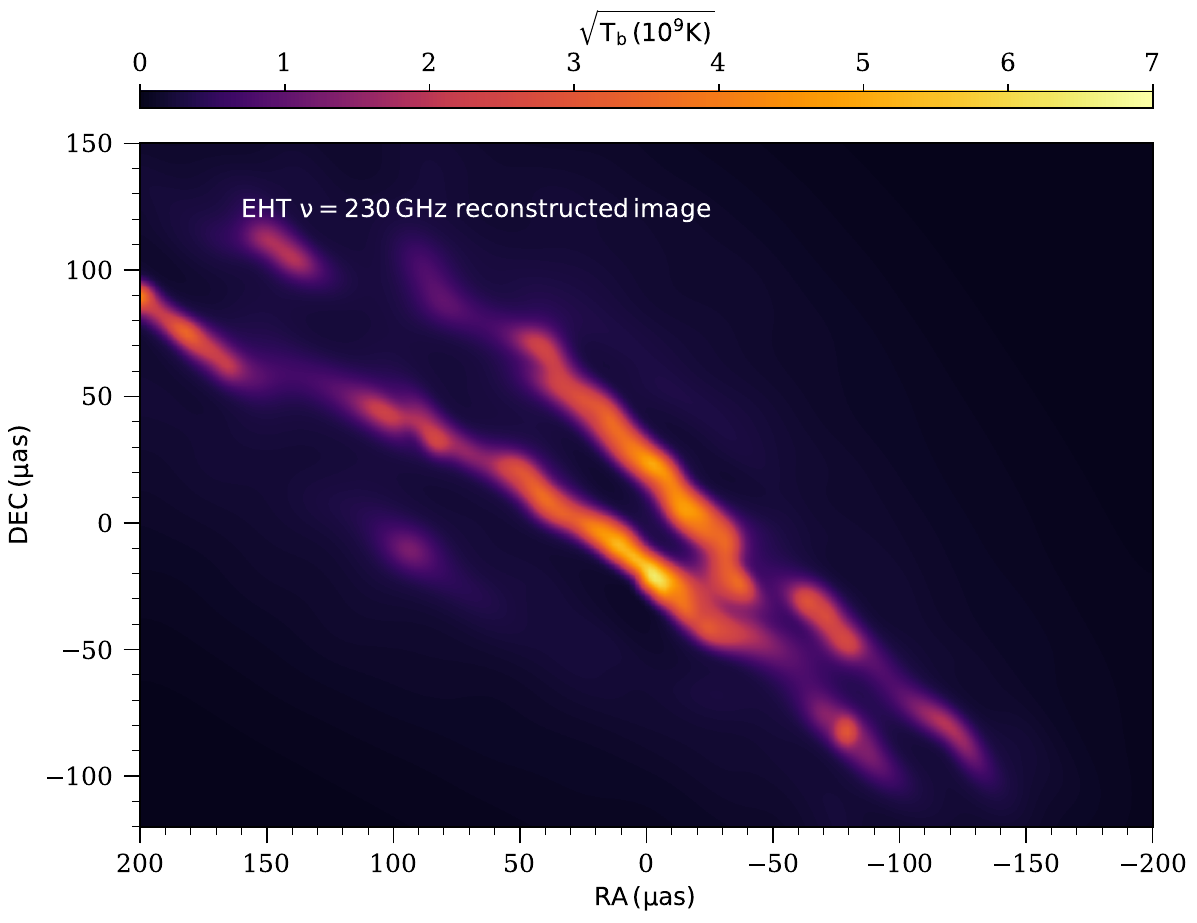}
    \caption{Reconstructed image from the EHT observations of Cen\,A \citep[][]{Janssen2021}}
    \label{fig:EHTimgdata}
\end{figure}

In order to adjust our model to the EHT observations, we need to adjust several free parameters. Namely, the mass of the black hole $M_{\rm BH}$, the mass unit $M$, the viewing angle $\vartheta$, the orientation of the source in the plane of the sky $\zeta$, the azimuthal orientation of the observer $\phi$, and the time of the used GRMHD snapshot. As mentioned earlier, the black hole mass is estimated to be $M_{\rm BH}=(5.5\pm3.0)\times10^7\,M_\odot$ \cite{Neumayer_2010} and low frequency VLBI observations suggest a viewing angle between $12^\circ\leq \vartheta \leq 45^\circ$ \citep{Mueller2014}. However, in order not to bias our modelling, we use a flat prior for the viewing angle from $0^\circ$ to $90^\circ$ and require at least a jet-to-counter-jet ratio of $R\geq 3$. Besides these estimates, the visibility of the EHT observations together with the jet-to-counter-jet ratio provides strong constraints on the model. In Fig.~\ref{fig:EHTimgdata} we present the EHT image of Cen\,A \citep[see][for details]{Janssen2021}. We include the modified System Equivalent Flux Densities (SEFDS) of the different telescopes according to Table 3 of \cite{EHT_M87_P3} and added 10\,\% systematic error and a noise floor of 0.01\,Jy during the MCMC runs to the EHT data similar to \cite{EHT_M87_P5}.

To apply our model to Cen\,A we coupled \texttt{IPOLE} with the Markow chain Monte Carlo (MCMC) code \texttt{emcee} \citep{Foreman2013}. The free parameters of our model and their prior distribution are listed in Table \ref{tab:prior}. 

\renewcommand{\arraystretch}{1.2}
\begin{table}[h]
    \centering
    \begin{tabular}{|l |c |l |}
        \hline
        parameter & range & prior type  \\
        \hline \hline
        $M_{\rm BH}$ (black hole mass)    & $(5.5\pm3.0)\times10^7\,M_\odot$   & Gaussian  \\
        $\vartheta$  (viewing angle)    & $0^\circ-90^\circ$   & flat \\
        $M$  (mass unit)   & $\left(10^{22}-10^{26}\right)\,\mathrm{g/cm^3}$   & flat \\
        $\eta$  (eDF anisotropy)   & $0.001-1$   & flat \\
        %$\phi$  (azimuth angle)   & $0^\circ-360^\circ$   & flat \\
        %$\zeta$  (rotation angle)   & $0^\circ-360^\circ$   & flat \\
        $t$ (time: GRMHD file)   & $(60-85)\,\mathrm{kM}$ & flat \\ 
        \hline
    \end{tabular}
    \caption{Parameters and their priors used for the MCMC runs.}
    \label{tab:prior}
\end{table}
\noindent
In addition to the visibility data from the EHT, we include constraints that we obtain a total flux of $\sim$1\,Jy in the image and jet-to-counter-jet ratio of $R\geq 2$. We initialize the MCMC runs with 500 walkers and 1000 steps to explore the parameter space listed in Table \ref{tab:prior} and use the position of the highest likelihood as initial for a second, refined MCMC run. The refined MCMC included 500 walkers with 5000 steps and converged after $\sim 1000$ steps. Notice that for each MCMC step, we perform an individual GRRT calculation (see Sect.~\ref{sec:GRRT}). During the radiative transport, we apply a magnetization cut-off of $\sigma_{\rm cut}=1$, which ensures to exclude the low-density spine affected by our floor model (see Sect.~\ref{sec:GRMHD}). Once an image is generated, we compute the jet-to-counter-jet ratio:
\begin{equation}\label{jet-to-counterjet R}
    R = {\int_{\mathrm{E}} \mathrm{d}x\mathrm{d}y\, I(x,y) }\,\bigg{/}{\int_\mathrm{W} \mathrm{d}x\mathrm{d}y\, I(x,y) },
\end{equation}
where $\{\rm E,W\}$ are the eastern and western half-planes, respectively. The condition chosen for separation is sufficient, since the optically-thick disk separates the jet and counter-jet wide enough in the W-E direction and is itself approximately canceled out in Eq. \eqref{jet-to-counterjet R}. 
If $R\geq2$, the complex visibility for synthetic EHT observations is generated via $V_{ij} =\int\hspace{-.2cm}\int I(x,y)e^{-2\pi \textrm{i}(ux+vy)}\,\textrm{d}x\,\textrm{d}y,$
\noindent
where $(i,j)$ are the baseline formed by the telescope pair $i,j$. $I(x,y)$ corresponds to the brightness distribution in the sky at the location $x,y$, and  $u,v$ indicates the projected baseline of the telescope pairs in Fourier space. From the generated complex visibilities $V_{ij}$, we compute the visibility amplitude $\left | V_{ij}\right|$ and the phase of closed antenna triangles.
MCMC minimizes logarithmic likelihood in the 5D-parameter space, where the likelihood function is derived from the sum of $\chi^2$ in visibility amplitudes $\chi_{\rm vis.\, amp.}^2$, and in closure phases, $\chi^2_{\rm cphase}$.

\section{Results}
\label{sec:results}

\subsection{Markow chain Monte Carlo method}\label{sec: results MCMC}

\begin{figure}[h!]
    \centering
    \includegraphics[width=\linewidth]{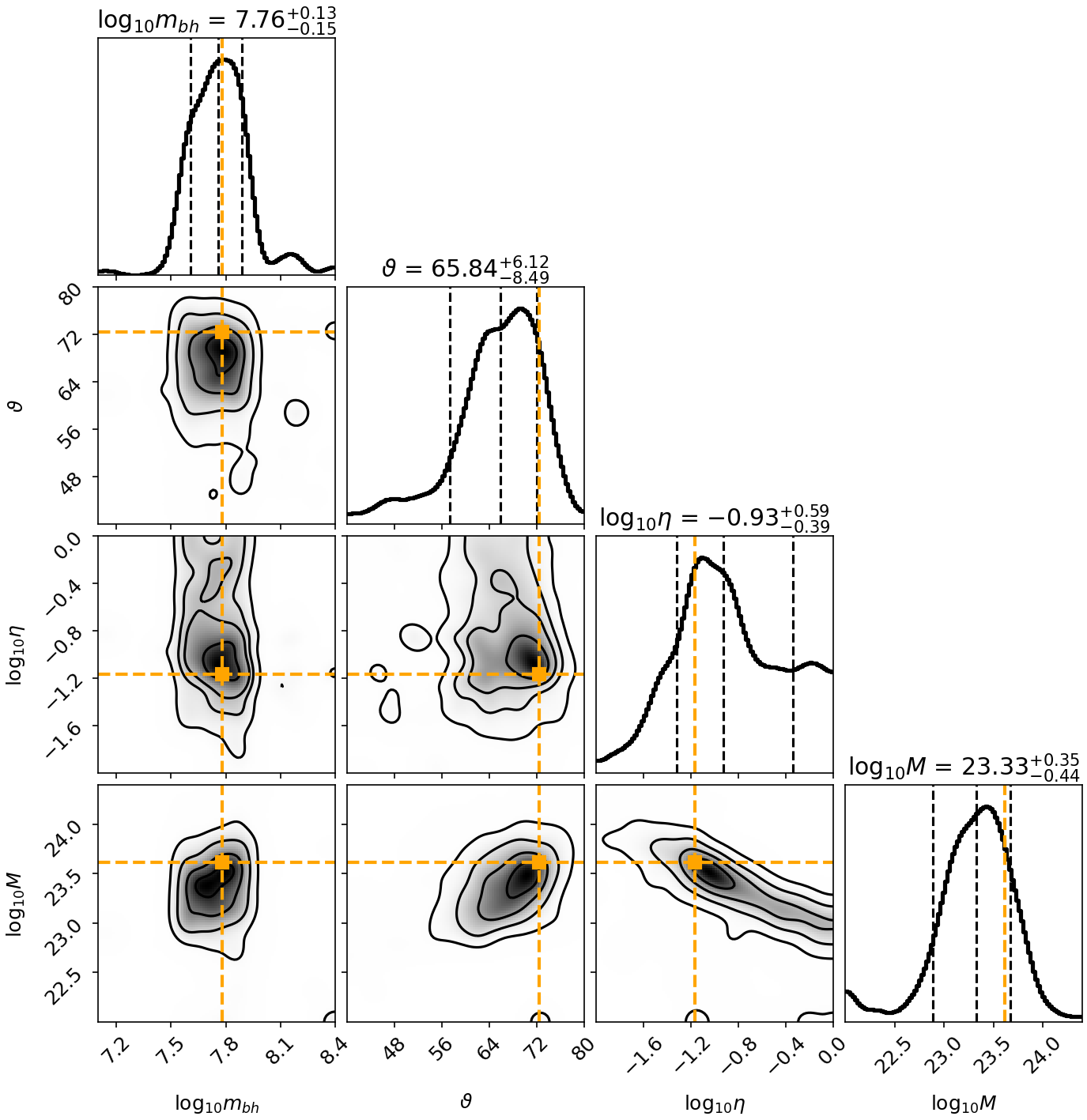}
    \caption{Posterior distribution of the black hole mass, $M_{\rm BH}$, viewing angle, $\vartheta$,  eDF anisotropy, $\eta$ and  mass unit, $M$. The black dashed lines correspond to mean and $1\sigma$ errors of the distributions, while the orange lines indicate the location of the model with the highest likelihood, i.e., best $\chi_{\rm tot}^2$.}
    \label{fig:posterior}
\end{figure}

In the following, we present the results of our modeling. In Fig. \ref{fig:posterior} we show the posterior distribution for the black hole mass, $M_{\rm{bh}}$, the viewing angle, $\vartheta$, the eDF anisotropy $\eta$, and the mass unit $M$. The model with the highest likelihood corresponding to the model with the lowest total $\chi^2_{\rm{tot}}$ is marked in orange, and the connected values are listed in Table~\ref{tab:mcmcres}

\begin{figure}[h!]
    \centering
    \includegraphics[width=\linewidth]{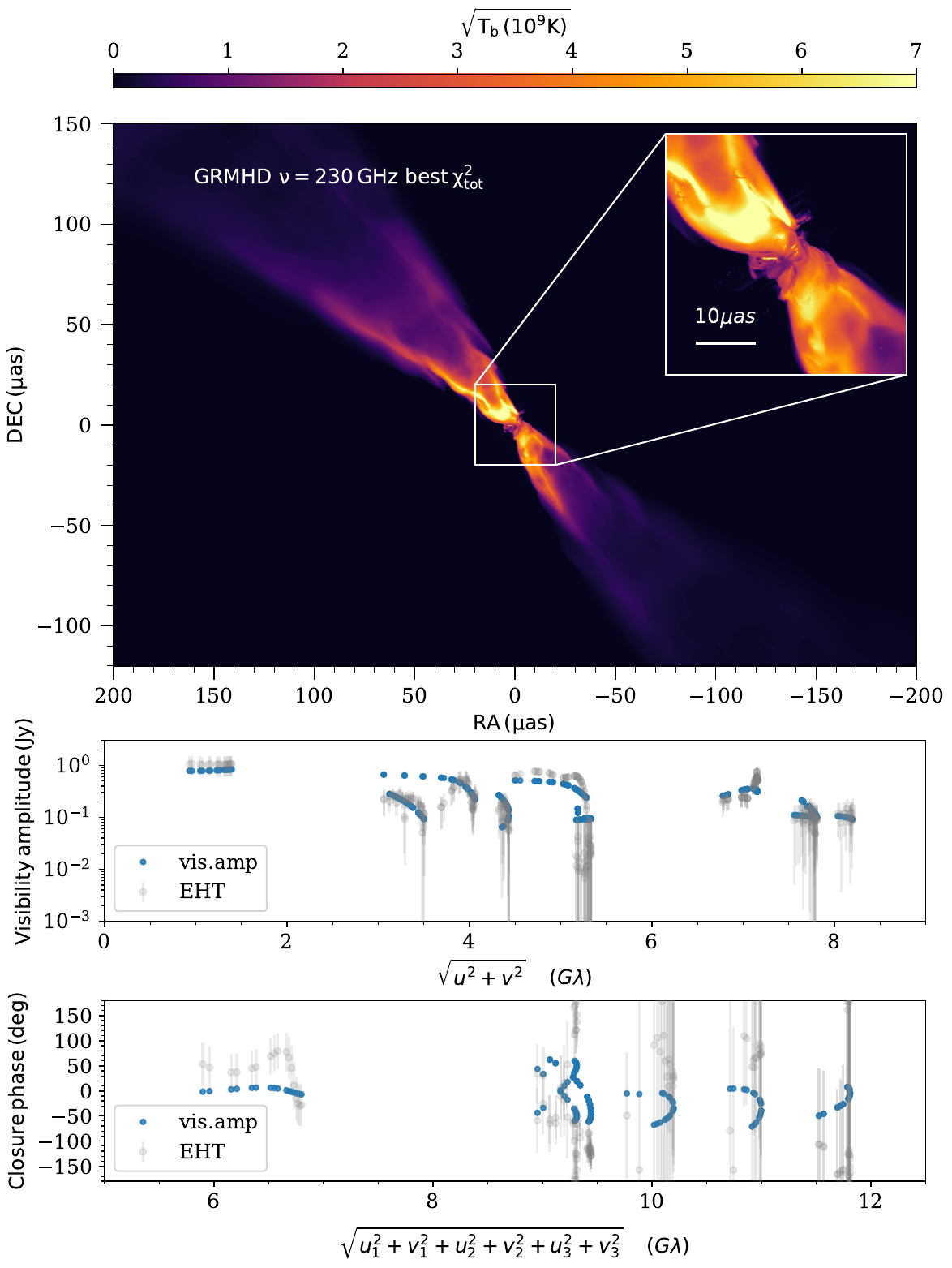}
    \caption{Result of the MCMC runs. The top panel shows a 230\,GHz GRRT image of Cen\,A computed using the maximum likelihood position of the MCMC run together with a zoom displaying the horizon scale structure. The middle panel reports on the visibility amplitudes of the EHT observation of Cen\,A \citep{Janssen2021} (gray) and our best GRRT image (blue). In the bottom panel, we report the closure phases for the EHT observations of Cen\,A \citep{Janssen2021} (gray) and our best GRRT image (blue).}
    \label{fig:MCMCres}
\end{figure}

\renewcommand{\arraystretch}{1.2}
\begin{table}[h]
    \centering
    \begin{tabular}{|l |l |}
        \hline
        parameter & highest likelihood \\
        \hline \hline
        $M_{\rm BH}$ (black hole mass)    & $6.01\times10^7\,M_\odot$ \\
        $\vartheta$  (viewing angle)    & $72.4^\circ$  \\
        $M$  (mass unit)   & $4.12\times10^{23}\,\mathrm{g/cm^3}$  \\
        $\eta$  (eDF anisotropy)   & $0.07$ \\

        $t$ (time GRMHD file)   & $64.7\,\mathrm{kM}$\\ 
        $\chi_{\rm vis.amp}^2,\,\,\chi_{\rm cphase}^2$                & 2.0, 1.8 \\ 
        \hline
    \end{tabular}
    \caption{Results of the MCMC runs for the models with the lowest $\chi^2$ on the visibility amplitude and with the lowest total $\chi^2$, i.e., on visibility amplitude and closure phase.} %\fg{add statistical uncertainties from MCMC?}}
    \label{tab:mcmcres}
\end{table}
The image obtained at 230\,GHz with a resolution of $0.2\,\mu\mathrm{as}$, together with the corresponding visibility amplitudes and closure phases evaluated at the maximum-likelihood position (see Table~\ref{tab:mcmcres}), is shown in Fig.~\ref{fig:MCMCres}. The image exhibits a pronounced edge-brightened structure in both the jet and the counter-jet. As in the reconstructed EHT image, the southern limb of the forward jet is brighter than the northern one. A central opacity gap is also present, though it is less pronounced than in the EHT reconstruction. The jet spine in our high-resolution GRRT image appears somewhat more filled than in the observations. A deeper comparison with the reconstructed EHT image of \citet{Janssen2021} would require performing an image reconstruction based on synthetic visibilities, which is beyond the scope of this work and defer to future studies. Overall, the visibilities computed from our GRRT image show good agreement with the EHT measurements, with the exception of a discrepancy in the range of 3.6–3.8\,$G\lambda$ (see middle and bottom panels). These are due to the substantial amplitude variations between and within VLBI scans caused by pointing problems that the Large Millimeter Telescope Alfonso Serrano (LMT) and South Pole Telescope (SPT) encountered in the 2017 EHT observation of Cen A, leading to high and variable gain errors (e.g., \cite{Janssen2021}).

In the following, we turn to the correlations obtained between the different parameters of the model. Three clear trends emerge between 
\begin{enumerate}
    \item the mass unit, $M$, and the black hole mass, $M_{\rm BH}$,
    \item the mass unit, $M$, and the viewing angle, $\vartheta$,
    \item and the mass unit, $M$, and the eDF anisotropy parameter, $\eta$.
\end{enumerate}

Before discussing these correlations, it is essential to recall how physical quantities are obtained from the dimensionless and non-radiative GRMHD simulations and how this scaling affects the synchrotron emissivity. We convert the simulation output to physical units by applying
\[
\rho_{\rm cgs} = m_{\rm unit}\,\rho_{\rm code}, \qquad
p_{\rm cgs} = m_{\rm unit}\,p_{\rm code}, \qquad
B_{\rm cgs} = \sqrt{m_{\rm unit}}\,B_{\rm code}.
\]
Because the electron number density scales as $n_{\rm e} \propto \rho_{\rm cgs}$ and the magnetic field scales as $B \propto m_{\rm unit}^{1/2}$, the synchrotron emissivity 
$j_\nu \propto n_{\rm e} B$ scales as $j_\nu \propto m_{\rm unit}^{3/2}$.
Given these scalings, we now interpret the correlations recovered in our parameter space exploration. \\

1) Increasing the black hole mass, $M_{\rm BH}$, changes the characteristic length scale of the simulation through the gravitational radius, $r_g = G M_{\rm BH}/c^2$. Because our GRRT calculations include radiative cooling, a larger $M_{\rm BH}$ shifts the synchrotron cooling break to lower frequencies (see Eqn.~\ref{eq:gbreak}). Since the emission at a fixed observing frequency scales as $(\nu/\nu_{\rm break})^{-1/2}$, a decrease in $\nu_{\rm break}$ reduces the observed flux density. To compensate for this drop and recover the target flux density, i.e., the zero-baseline flux density, the model requires a higher mass unit, $m_{\rm unit}$, which increases the synchrotron emissivity as $j_\nu \propto m_{\rm unit}^{3/2}$.

2) Increasing the viewing angle, $\vartheta$, reduces the Doppler boosting of the emission. The Doppler factor is 
$\delta = \frac{1}{\gamma (1 - \beta \cos\vartheta)}$, where $\beta = v/c$ and $\gamma = (1-\beta^2)^{-1/2}$ are the bulk velocity and Lorentz factor of the emitting plasma. For optically thin synchrotron radiation, the emissivity transforms as $j_\nu = \delta^2 j'_{\nu'}$, with $\nu = \delta\, \nu'$, where primed quantities refer to the plasma rest frame. As $\vartheta$ increases, the Doppler factor decreases, reducing the observed emissivity. To maintain the desired flux level, this reduction must be offset by increasing $m_{\rm unit}$, which enhances the intrinsic emissivity according to $j_\nu \propto m_{\rm unit}^{3/2}$.

3) Increasing the anisotropy of the electron distribution (i.e., decreasing $\eta$ towards $0$) enhances the emission from particles with small pitch angles, i.e., those moving more closely aligned with the magnetic field, while suppressing emission from particles with larger pitch angles (see Eqs.~\ref{eqn:anisoja}--\ref{eqn:peta} and Fig.~\ref{fig:aniso}). In practice, this shifts the brightness from the jet spine toward the jet sheath and reduces the total flux. To reach the required target flux, the model must again increase the mass unit, $m_{\rm unit}$, explaining the recovered correlation between $m_{\rm unit}$ and the distribution anisotropy.

\subsection{Dynamical and spectro-polarimetric behavior of the Cen\,A jet model}

In the following section, we explore the spectral and dynamical behavior of our jet model based on the results of our MCMC model and provide testable observable key signatures for current and future ground and space-based telescopes. We limit our modeling on the GRMHD simulations between $60\,\mathrm{kM} < t < 85\,\mathrm{kM}$. Using the extracted black hole mass of $6.01\times10^7\,M_\odot$ the characteristic time scale $t_g=GM_{\rm BH}/c^3=295\,\rm{s}$. Thus, our selected time range covers roughly 3 months in the evolution of Cen\,A. 

\subsubsection{Jet width, opening angle and limb-brightening}

\begin{figure}[h!]
    \centering
    \includegraphics[width=\linewidth]{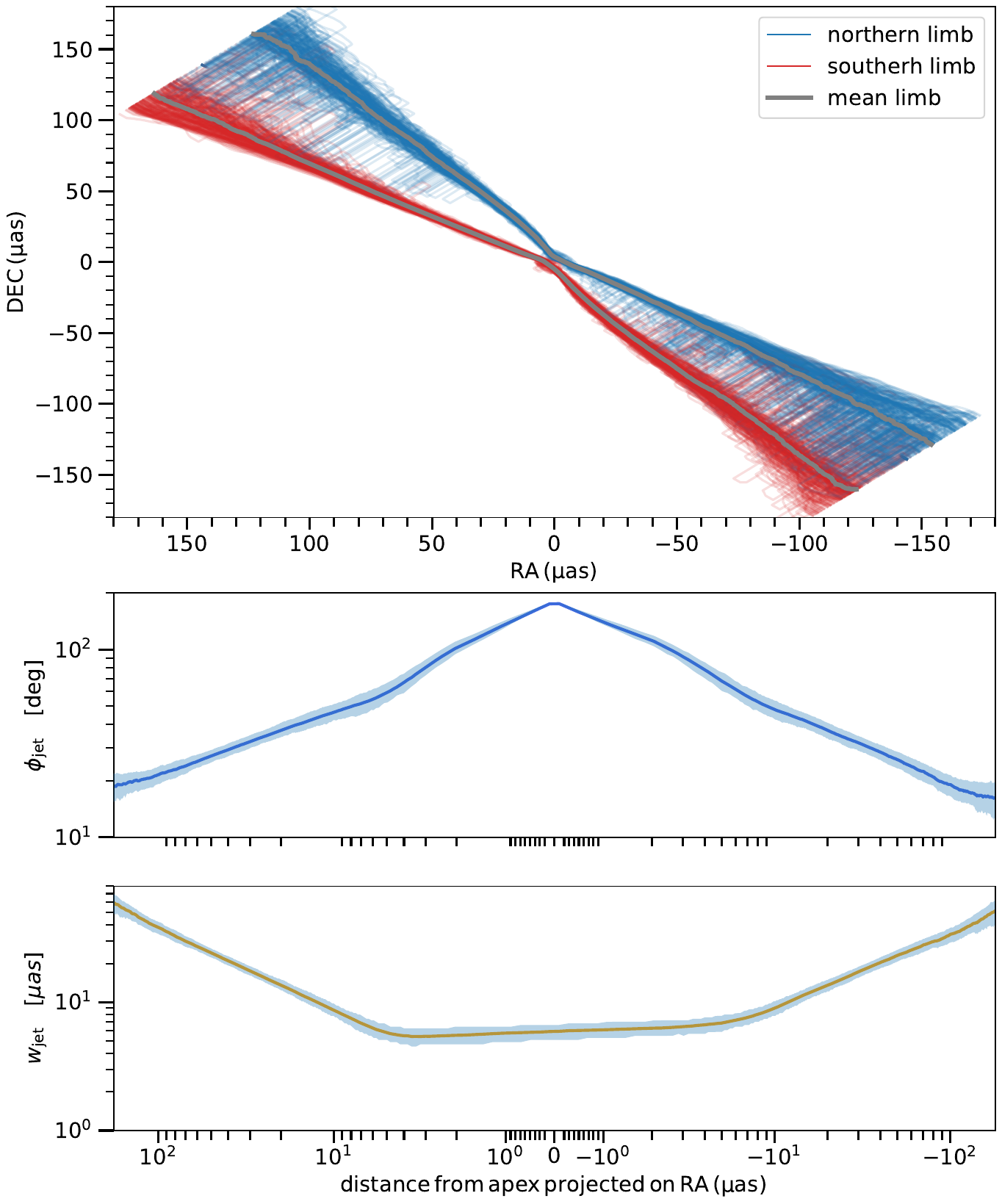}
    \caption{Top panel: Temporal evolution of the jets' northern and southern limb at 230\,GHz for $r<200\,\mu as$ including the mean position of the jet limbs (gray lines). Middle panel: Jet opening angle computed from the limb positions including 1$\sigma$ uncertainty band. Bottom panel: Jet width computed from the limb positions including 1$\sigma$ uncertainty band. Notice the different scaling of the axis: linear in the top panel and double logarithmic in the middle and bottom panels.}
    \label{fig:ridgeline}
\end{figure}

In order to explore the dynamics and structure of our Cen\,A model, we extract the limb location and compute the corresponding jet width, $w_{\rm jet}$, and opening angle $\phi_{\rm jet}$. For the location of the jet limbs, we rotate the jet to align its jet axis with the x-direction. In the next step, we slice the jet perpendicular to the jet axis (x-axis) and locate the outermost peaks in the transversal flux distribution. In order to focus on the most dominant one, we apply a Gaussian filter with a distance-dependent standard deviation. We model the distance-dependent standard deviation via a sigmoid function with a minimum value of $\sigma=2$ and a maximum value of $\sigma=10$ reached at a distance of 100 $\mu as$ from the jet apex. In addition, we enforce that the distance between the found peaks is at least $2\,\mu$as and increases with distance. This procedure ensures a reliable and robust detection of the jet limbs throughout our simulation. In the upper panel of Fig.~\ref{fig:ridgeline}, we present our results on the limb detection. The blue and red lines correspond to the individual limb location for the explored time span, and the gray lines indicate the mean position. Notice that for distance $>100\,\mu$as ($>50\,\mu$as) in the forward (counter-jet) from the jet apex, the flux has significantly dropped, leading to shallower limb-brightening. This leads to some variations in the location of the detected limbs (see the top panel in Fig.~\ref{fig:ridgeline}). Based on the limb positions, we compute the connected jet opening angle (second panel in Fig.~\ref{fig:ridgeline}) and the jet width (third panel in Fig.~\ref{fig:ridgeline}).

In order to quantify the jet collimation profile, we fitted a power-law to the jet width. Motivated by the profile of the extracted jet width, $w_{\rm jet}$, (see bottom panel in Fig.~\ref{fig:ridgeline}), we utilize a power-law model including a plateau for small distances from the jet apex in form of:
\begin{equation}
    w_{\rm jet}(z)=w_{\rm jet,0}\left[1+\left(\left |z/z_0\right|\right)^{k_1}\right]^{k_2}
\label{eq:wfit}
\end{equation}
In the above equation $w_{\rm jet,0}$ corresponds to the base width of the jet, $z_0$, to the transition distance i.e., the location of the start of the power-law expansion, $z$ to the distance along the jet\footnote{we perform the fit for jet and counter jet separately} and $k_{1,2}$ to the expansion exponents. Notice that for large distances from the apex, the jet width is given by $\propto |z|^{k_1 k_2}$. Applying Eq.~\ref{eq:wfit} to our extracted jet width, we obtain the following results:

\renewcommand{\arraystretch}{1.2}
\begin{table}[h]
    \centering
    \begin{tabular}{|l |l |l |}
        \hline
        parameter & forward jet & counter jet \\
        \hline \hline
        $w_{\rm jet,0}$  & $6.15\pm0.24$ & $5.48\pm0.24$ \\
        $z_0$            & $5.29\pm 0.36$ & $5.32\pm 0.35$\\
        $k_1$      & $3.22\pm 0.98$ &  $4.13\pm1.51$  \\
        $k_2$      & $0.18\pm0.06$ & $0.16\pm0.06$  \\
        $k_1k_2$   & $0.58\pm0.25$ & $0.67\pm0.35$   \\
        \hline
    \end{tabular}
    \caption{Results of the jet width fitting}
    \label{tab:jwidth}
\end{table}

\subsection{Variability of jet limbs and Kelvin--Helmholtz instability}

\begin{figure}[h!]
    \centering
    \includegraphics[width=\linewidth]{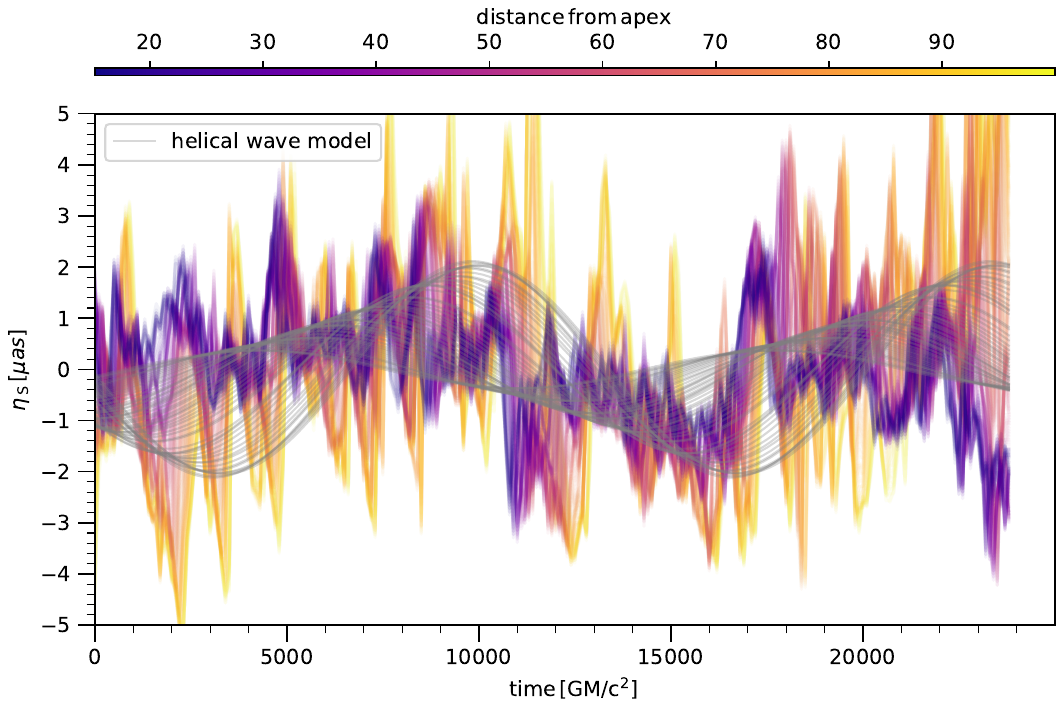}
    \caption{Temporal evolution of $\eta_S$ along the jet with over-plotted helical wave model.}
    \label{fig:helicalwave}
\end{figure}

Kelvin--Helmholtz (KH) instabilities are expected at velocity-shear interfaces in relativistic jets, such as the northern and southern limbs in our images. In the linear regime, relativistic (magneto)hydrodynamic dispersion relations predict spatial and temporal growth rates of surface modes. To compare theory with observations, we compute these rates for the forward and counter-jet limbs\footnote{forward jet points towards positive RA direction while the counter-jet towards negative RA direction, see also Fig.~\ref{fig:MCMCres}}. For each limb $L \in \{N,S\}$ we define the time-averaged interface $\bar{y}_L(x)$ and measure the transverse perturbations via:
\begin{equation}
\eta_L(x,t) = y_L(x,t) - \bar{y}_L(x).
\end{equation}
We fit the spatial growth rates using the RMS displacement envelope using $\langle \eta_L^2 \rangle^{1/2} \propto e^{\alpha_L x}$. For the northern limb, we measure a growth rate of $(1.58\pm0.02) \times10^{-2}\,[1/\mu as]$ in the forward jet and a similar value of $(1.55\pm0.02 )\times10^{-2}\,[1/\mu as]$ in the counter-jet. Similarly, the values for the southern limb in forward and counter jets are $(1.48\pm0.01 )\times10^{-2}\,[1/\mu as]$ and $(1.49\pm0.01 )\times10^{-2}\,[1/\mu as]$.
Our temporal analysis shows bounded oscillations rather than exponential growth, consistent with long-lived helical perturbations in magnetized flows. We decompose the oscillations into a slow, global helical mode and higher-frequency fluctuations. The slow mode could represent rigid rotation of field lines anchored to the central engine, while higher-frequency components maybe attributed to internal MHD body modes. We extract characteristic frequencies using a global fit across space and time. After detrending, the time series at each position is modeled as
\begin{equation}
y_j(t) = a_j \cos(\omega t) + b_j \sin(\omega t) = A_j \cos(\omega t + \phi_j),
\end{equation}
with coefficients $(a_j,b_j)$ determined by linear least squares, minimizing
\begin{equation}
\chi^2(\omega) = \sum_{j=1}^{N_x} \sum_{i=1}^{N_t} \big[y_j(t_i) - a_j \cos(\omega t_i) - b_j \sin(\omega t_i)\big]^2.
\end{equation}
We compute the uncertainty in the extracted frequency by the width of the $\chi^2(\omega)$ minimum. 
In Fig.~\ref{fig:helicalwave}, we present the results of our modeling for the southern limb in the forward jet. The top panel shows the temporal variation of the transverse perturbation for different positions along the southern limb (color-coded), and the gray lines correspond to the helical wave model. The amplitude of the helical wave increases with distance $z$ from the apex as $A_j\propto z^{0.02}$ while the phase $\phi_j\propto z^{-0.03}$ decreases.

%The amplitude of the helical wave, $A_j$, along the limb is presented in the bottom left panel, while the changes in the phase $\phi_j$ along the limb can be seen in the bottom right panel.

The best-fitting global frequencies are $\omega \simeq (3.5\pm1.1)\times10^{-4}$ (north) and $(4.2\pm1.2)\times10^{-4}$ (south) for the forward jet and $\omega \simeq (2.4\pm1.0)\times10^{-4}$ for both counter-jet limbs. The differences between jet and counter-jet arise from Doppler effects:
\begin{equation}
\omega_{\rm obs} = \delta \, \omega_{\rm int}, \quad \delta=[\Gamma(1-\beta\cos\theta_{\rm eff})]^{-1},
\end{equation}
with $\theta_{\rm eff} \approx \theta_{\rm view} \pm \theta_{\rm open}/2$. Correcting for Doppler shifts aligns the intrinsic frequencies of all limbs.

The observed slow helical frequencies are much lower than the expected angular velocity of field lines anchored to a Kerr black hole ($\Omega_F \sim \Omega_H/2=0.175c^3/GM$ for $a=0.94$), indicating that they trace large-scale helical patterns downstream rather than black hole spin. In relativistic MHD, these are magnetically dominated surface or body waves with phase speed $\omega_{\rm pat} \simeq k v_{\rm ph}$, stabilized by toroidal magnetic tension. The absence of exponential temporal growth is consistent with the suppression of KH instabilities in magnetized jets \citep[see, e.g.,][]{Hardee2007,Mizuno2007,Sinnis2023}.

Higher-frequency oscillations, superposed on the slow helical mode, are consistent with internal MHD body modes and potentially reflect field-line rotation. Notice that in magnetically dominated flows ,current-driven instabilities grow faster than KH instabilities. However, the current temporal resolution does not allow robust extraction of these high-frequency signals. The slow oscillation provides insight into downstream jet stability and magnetization, while direct measurement of black hole spin from time-domain variability requires higher-cadence data.

\subsubsection{Light-curve and broad-band spectrum}

\begin{figure}[h!]
    \centering
    \includegraphics[width=\linewidth]{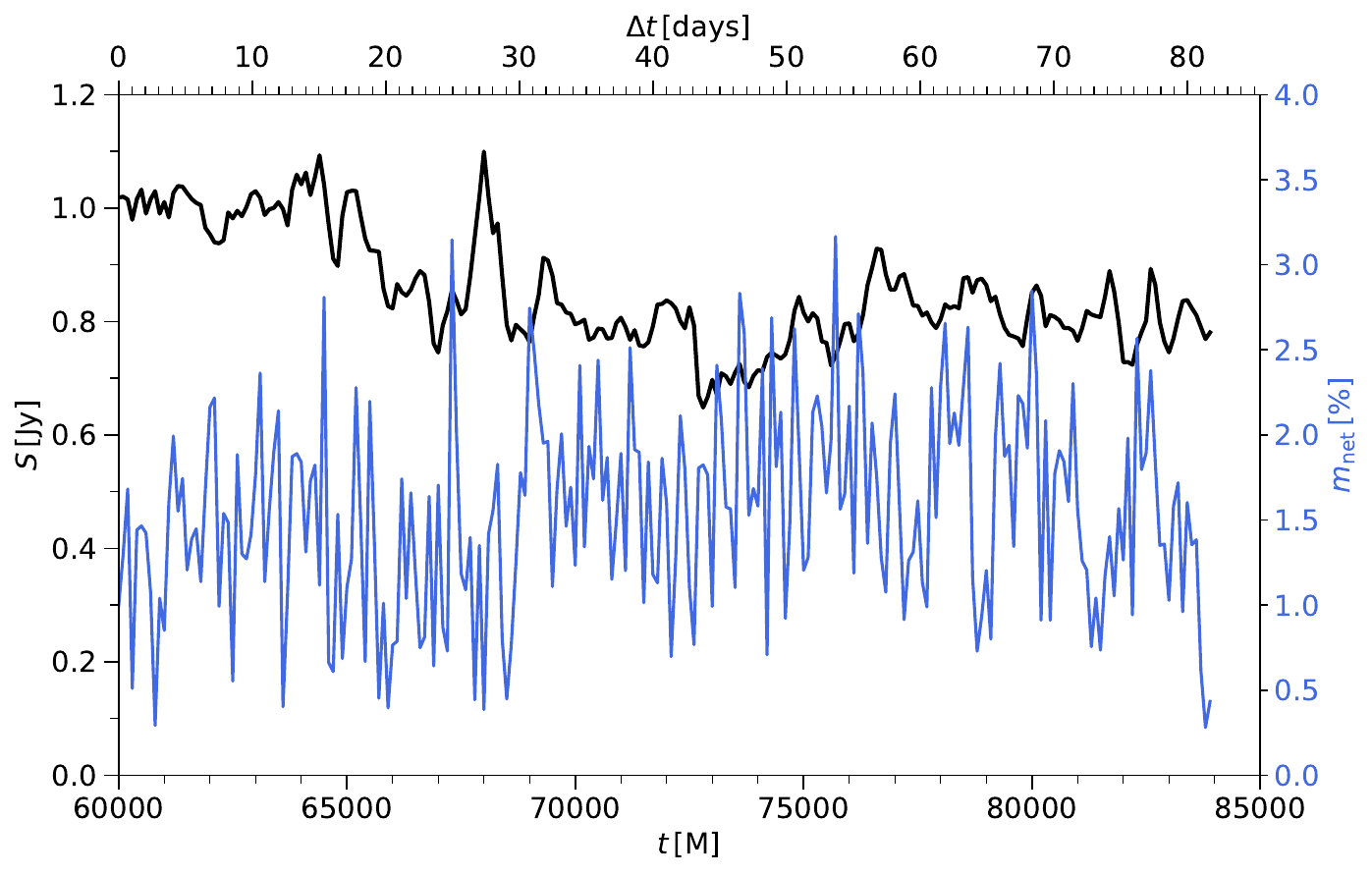}
    \caption{Temporal evolution of the 230\,GHz flux density (black) and unresolved linear polarization fraction (blue) for a time interval of 85 days (25kM).}
    \label{fig:230flux}
\end{figure}

In Figure~\ref{fig:230flux} we present the 230\,GHz light curve and the unresolved (image-integrated) linear polarization fraction in percent defined as:
\begin{equation}
    m_{\rm net}=\frac{\sqrt{\left(\Sigma_i Q_i\right)^2+\left(\Sigma_i U_i\right)^2}}{\Sigma_i I_i},
\end{equation}
where I, ,Q and U correspond to Stokes parameters and the subscript $i$ indicates the individual pixels of the images.

The average flux density at 230\,GHz is $\langle S_{230} \rangle = 0.85 \pm 0.10\,\mathrm{Jy}$, and the mean unresolved linear polarization fraction is $\langle m_{\rm net} \rangle = 1.5 \pm 0.5\%$. The 230\,GHz light curve shows several modest flares and one prominent, isolated flare at $t = 68\,\mathrm{kM}$. Overall, the variability can be described as two quasi-stationary plateaus: an early phase with a mean flux of $\sim 1\,\mathrm{Jy}$ for $t < 68\,\mathrm{kM}$, and a later phase around $\sim 0.8\,\mathrm{Jy}$ for $t > 75\,\mathrm{kM}$. These two regimes are connected by a smooth transitional decline, interrupted only by the strong flare at $t = 68\,\mathrm{kM}$. Our lower value of the linear polarization fraction is consistent with the upper limit reported for Centaurus~A from ALMA-only observations, $\langle m_{\rm net} \rangle = 0.07 \pm 0.03\%$ \citep{Goddi2021}. It is important to note that the ALMA observations cover a field of view of several tens of arcseconds, whereas our simulations only probe the innermost $\sim 400\,\mu\mathrm{as}$. Consequently, substantial depolarization from large-scale structures that are not included in our simulated domain is expected, naturally leading to lower observed polarization fractions in the ALMA data.

\begin{figure}[h!]
    \centering
    \includegraphics[width=\linewidth]{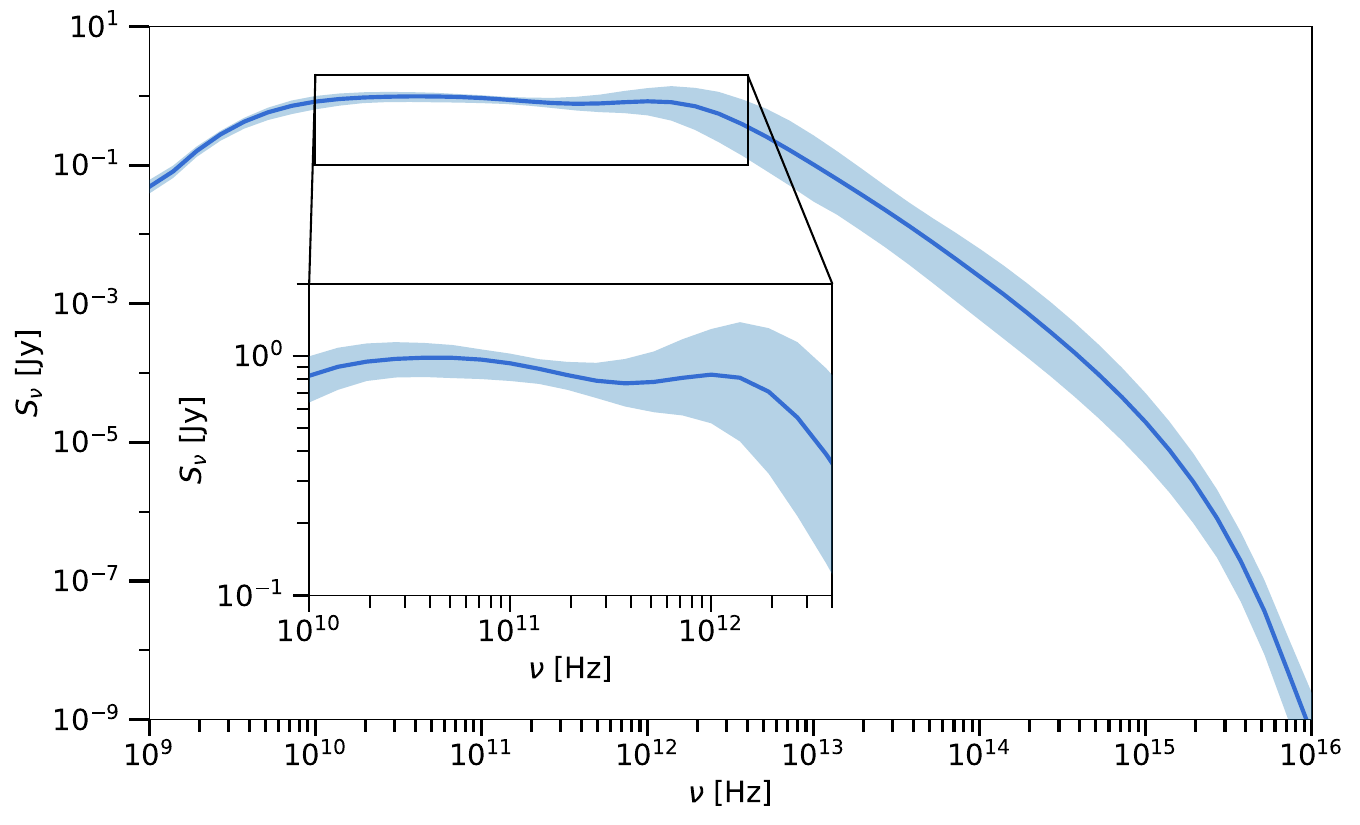}
    \caption{Temporal evolution of the broad-band spectrum from $10^{9}$\,Hz to $10^{16}$\, Hz with a time interval of 85 days (25kM). The solid line corresponds to the mean, and the light blue band indicates the $1\sigma$ region. The inlet provides a zoom into the 10\,GHz to 3\,THz regime.}
    \label{fig:spectrum}
\end{figure}

\begin{figure*}[t!]
    \centering
    \includegraphics[width=0.99\linewidth]{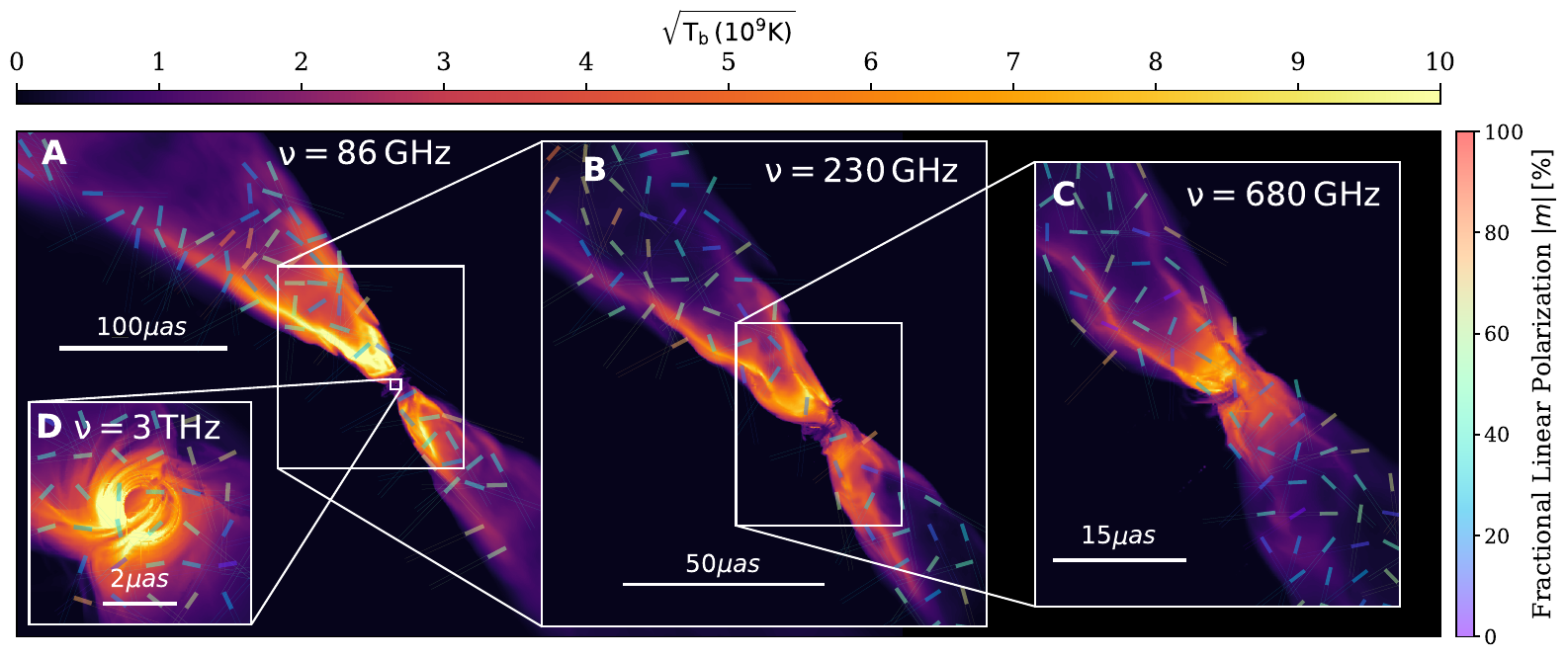}
    \caption{Multi-frequency images of our best model for Cen\,A from 86\,GHz to 3\,THz. The over-plotted bars indicate the EVPAs with color-coded fractional linear polarization. The jet-dominated non-thermal emission originates from an optically thick, thermal core which turns optically thin with increasing frequency, revealing the accretion disk as well as the black hole shadow at 3\,THz.}
    \label{fig:CenAmultifreq}
\end{figure*}

The broad band spectrum between $10^{9}$\,Hz and $10^{16}$\,Hz over a time span of $25\,\mathrm{kM}$ ($\simeq 3$ months) is displayed in Fig.~\ref{fig:spectrum} 
The spectrum displays three characteristic regimes. 
At low frequencies ($\nu \lesssim 10^{10}$\,Hz), the emission is optically thick and follows a positive spectral index ($S_\nu\propto v^{+\alpha})$, as expected for self–absorbed synchrotron emission from the compact jet base and the innermost accretion flow. 
At intermediate frequencies ($10^{10} \lesssim \nu/{\rm Hz} \lesssim 10^{12}$) the spectrum becomes flat. Such flat radio-to-mm spectra are a well-known property of stratified, partially self-absorbed jets \citep[see, e.g.,][]{Blandford1979}. In our case, this emission originates primarily from the jet, where we model the electron population via a $\kappa$-distribution (see Eq. \ref{eq: kappa-eDF}). The presence of a non-thermal power-law tail of the kappa eDF in the jet with varying slope (see Eq. \ref{eq:kappavalue}) enhances the optically thin contribution from different jet segments whose synchrotron turnover frequencies span this range, and the superposition of these components produces the observed flat spectrum. In addition, the radiative cooling included in the emission calculations leads to additional steepening of the individual spectra and shifts the turnover frequencies towards smaller values. In order to compare the spectral behavior of our model with the ALMA-only observations of Cen\,A, we computed the spectral index between 213\,GHz and 229\,GHz mode. We obtained a mean value of $\langle \alpha\rangle = -0.16$, which is in agreement with the reported value of $\langle \alpha\rangle=-0.197 \pm 0.038$ \citep{Goddi2021}.

At higher frequencies ($\nu \gtrsim 10^{12}$\,Hz), the emission becomes optically thin, and the spectrum transitions into a steeper power law dominated by the non-thermal tail of the jet’s $\kappa$-distribution and the exponential decay of the thermal emissivities of the accretion disk. Beyond $\sim10^{15}$\,Hz the spectrum exhibits a pronounced exponential suppression. This high-frequency cut-off is a consequence of synchrotron cooling in the jet sheath and spine, which depletes the highest-energy electrons and introduces a cooling break in the particle energy distribution. The location and amplitude of this break vary over time, reflecting changes in underlying plasma, including variations in the magnetic energy density, particle density, and acceleration efficiency (see Eq. \ref{eq:kappaeff}) within the jet launching region.

\subsubsection{Multi-frequency and spectral index images}
 
Based on the spectral shape, especially its variability we conclude that at frequencies larger than $\nu>300$\,GHz will allow us to study the flaring and disk-jet coupling best while the horizon structure including the black hole shadow should become visible at frequencies $\nu>1$\,THz in agreement with the analytic estimates in \cite{Janssen2021}. 

In Figure~\ref{fig:CenAmultifreq} we present the synthetic polarized intensity images of our Cen~A model at four representative frequencies: 86\,GHz (panel A), 230\,GHz (panel B), 680\,GHz (panel C), and 3\,THz (panel D). The images are displayed with color maps corresponding to the square root of the brightness temperature 
\(\sqrt{T_b/(10^{9}\,\mathrm{K})}\) with over-plotted ticks indicating the electric vector position angles (EVPAs) and the color-coded linear polarization fraction.
At 86\,GHz (panel A), our model extends up to 400\,$\mu as$ with a highly pronounced jet brightening in the forward jet and a clear absorption gap between the jet and counter jet. At the EHT observing frequency of 230\,GHz (panel B), the image captures the extended jet and counter-jet on scales of several tens of microarcseconds. The brightness temperature distribution reveals a well-collimated jet with prominent limb brightening, as expected for our anisotropic particle acceleration model. The polarization vectors show a complex pattern with EVPAs aligned both parallel and perpendicular to the jet axis in different regions. This likely reflects a mixture of poloidal and toroidal magnetic field components as well as possible Faraday effects in the dense plasma near the core and rotation due to relativistic aberration \citep[see e.g.,][]{Lyutikov2005}.

At 680\,GHz (panel C), the image zooms into the inner jet within a scale of approximately 50 micro-arcseconds. The brightness temperature in the core region increases compared to 230\,GHz, indicating that the emission is dominated by more compact regions closer to the black hole (see also broad-band spectrum in Fig.~\ref{fig:spectrum}). The polarization pattern becomes more ordered, with EVPAs predominantly aligned along the outer ridges of the jet, consistent with a helical magnetic field geometry. This enhanced order arises from reduced Faraday depolarization and optical depth effects at higher frequencies, allowing a clearer view of the intrinsic magnetic structure in the jet and jet launching region \citep[see e.g.,][in the case RMHD jet simulations]{Hu2025}

\begin{figure}[h!]
    \centering
    \includegraphics[width=\linewidth]{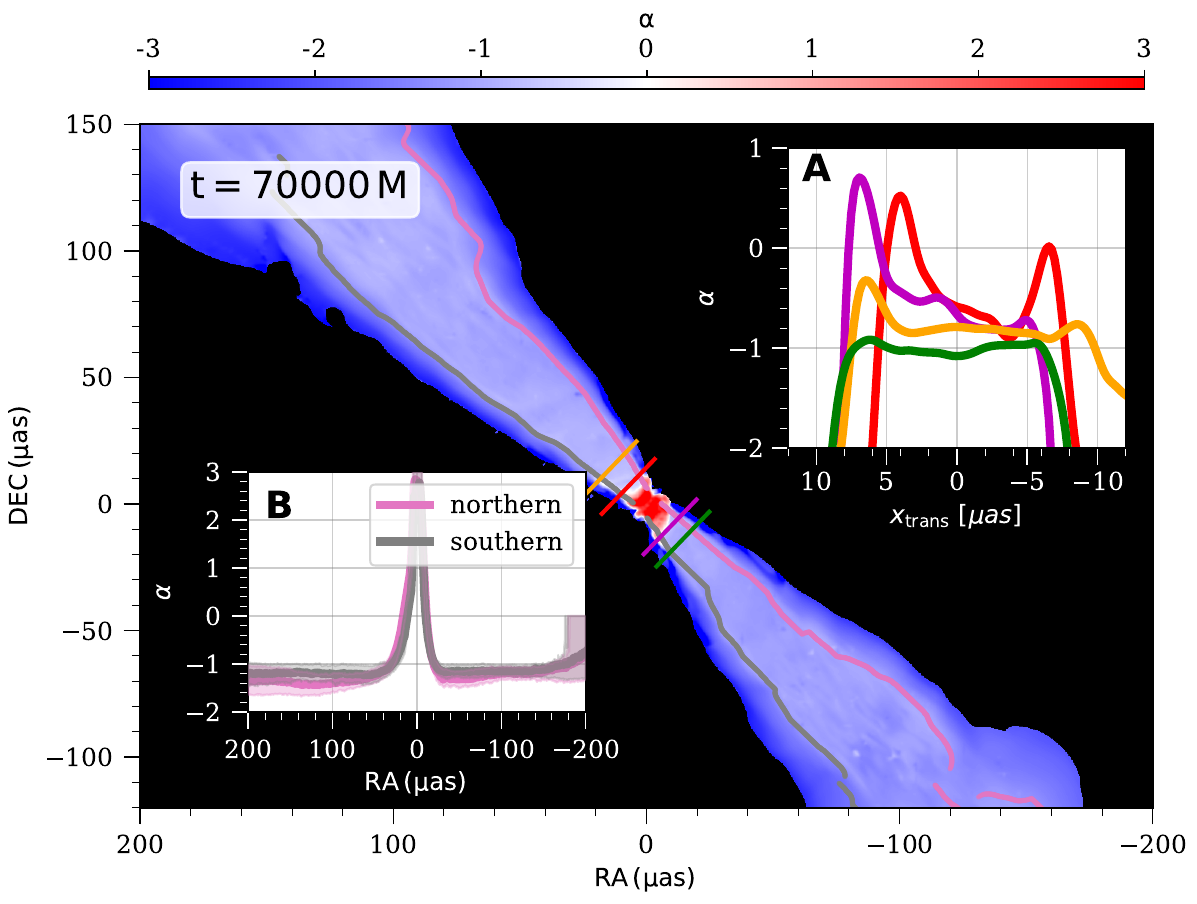}
    \caption{Spatial distribution of the Spectral index, $\alpha$ between 230\,GHz and 345\,GHz at $t=70\,\mathrm{kM}$. Inlet A shows the transversal spectral index profile for four selected locations and inlet B the average spectral index with 1$\sigma$ variation along the northern (pink) and southern (gray) jet limb (see top panel for location) computed between $60\,\mathrm{kM} < t < 85\,\mathrm{kM}$.}
    \label{fig:spix}
\end{figure}

\begin{figure*}[h!]
    \centering
    \includegraphics[width=0.99\linewidth]{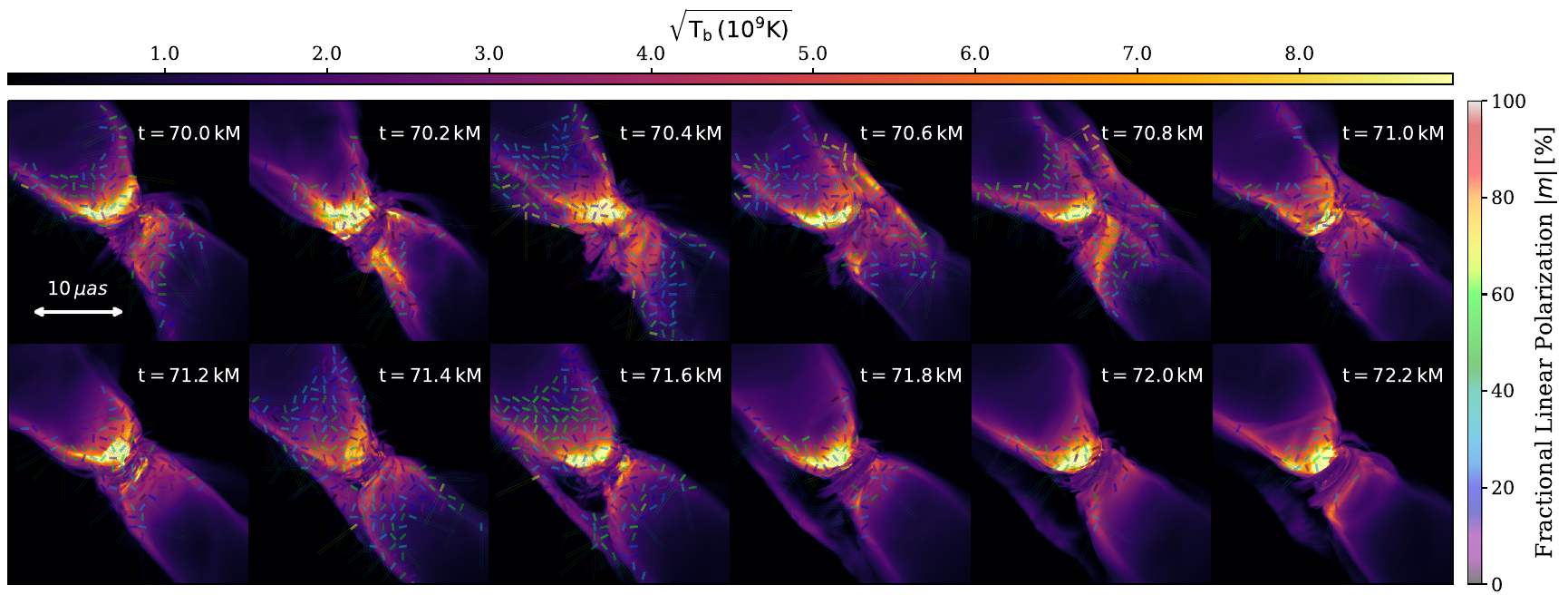}
    \caption{Evolution of MAD flux tubes indicated by flux streams parallel to the jet direction orbiting the horizon scale structure of Cen\,A during a flaring event at 680\,GHz. The over-plotted bars indicate the EVPAs with color-coded fractional linear polarization.}
    \label{fig:fluxtubes}
\end{figure*}

The highest frequency image at 3\,THz (panel D) resolves the innermost accretion flow and jet launching region on scales of a $\sim$few micro-arcseconds. The brightness temperature peaks sharply, reflecting emission from the hottest plasma closest to the event horizon, i.e., the inner regions/edges of the accretion disk, clearly revealing the black hole shadow. The bright ridge in front of the black hole shadow provides strong constraints on the orientation of the black hole towards the observer. The EVPA vectors reveal a distinctive spiral or azimuthal pattern around the black hole, suggestive of a predominantly toroidal magnetic field. Notice the alignment of the EVPAs along the flux arcs connecting the black hole with the jets forming the jet base.

In Fig.~\ref{fig:spix} we provide the spectral index computed between 230\,GHz and 345\,GHz. The horizon scale region is optically thick, followed by a sharp transition to optically thin emission. For $r<30\,\mu as$ from the black hole, the transversal spectral index profiles show two peaks located at the northern and southern limbs. These peaks are optically thick and smoothly transition to optically thin at larger distances. This behavior indicates on-going particle acceleration along the jet limbs within $r<20\,\mu as$. (see Fig. \ref{fig:spix} and inlet A). Along the limbs, the spectral index reaches a stable value of $\alpha\sim-1.3$ after $r>30\,\mu as$ (see inlet B in Fig.~\ref{fig:spix}).

\subsubsection{Horizon structure and flux tubes}

One of the defining characteristics of MAD simulations is the frequent formation of magnetic flux tubes near the central black hole. These structures are subsequently advected into the accretion disk, where they co-rotate with the flow and are progressively sheared apart \citep{Porth2021}. Detecting and tracking such flux tubes would therefore (i) provide direct observational confirmation of the MAD accretion state and (ii) yield valuable insights into the dynamics of the accretion flow, including the velocity profile of the disk.

Probing the near-horizon region and the associated orbiting flux tubes requires observations at high frequencies, $\nu > 345\,\mathrm{GHz}$, in order to mitigate synchrotron self-absorption caused by the dense accretion disk obscuring the central black hole (see Fig.~\ref{fig:CenAmultifreq}). In addition, very high angular resolution, $\Theta \sim 5\,\mu\mathrm{as}$, is necessary to resolve near-horizon structures, including individual flux tubes. These observational requirements are expected to be met by future space-based interferometers such as \texttt{BHEX} \citep{Johnson2024} and \texttt{SHARP}.

In the following, we outline potential observable signatures of orbiting flux tubes. Figure~\ref{fig:fluxtubes} presents a sequence of near-horizon-scale images at 680\,GHz, capturing a flux eruption and the subsequent formation of an orbiting flux tube. The flux tube appears as an elongated emission feature aligned approximately parallel to the main jet axis. During its temporal evolution, the structure is initially displaced radially outward, followed by inward motion, before temporarily disappearing as it moves behind the black hole and reappearing on the opposite side. This behavior reflects projection effects at our assumed inclination angle of $\vartheta = 72^\circ$.
\begin{figure}[h!]
    \centering
    \includegraphics[width=\linewidth]{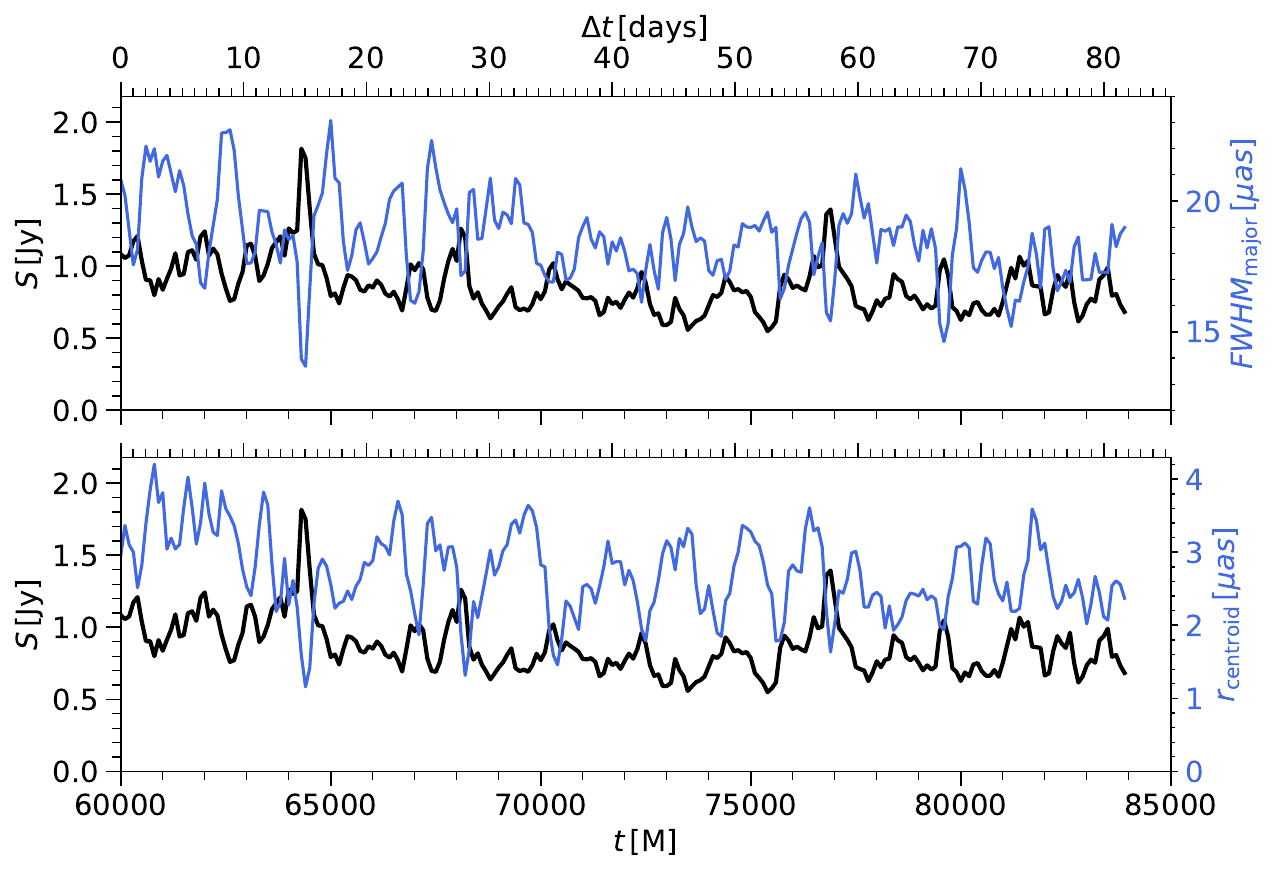}
    \caption{Temporal evolution of the flux density at 680 GHz over-plotted with changes in the image size (top) and position of the image centroid (bottom). Notice the anti-correlation of the image size and image centroid with the flux density. The parameters are computed between $60\,\mathrm{kM} < t < 85\,\mathrm{kM}$ covering roughly 3 months in the observer's frame.}%\fg{Add indicating areas or lines for MAD-states}}
    \label{fig:unresolved_signatures}
\end{figure}
The formation of flux tubes is found to be slightly anti-correlated with flaring activity, with flux tubes predominantly emerging during the decay phase of flares, in agreement with the findings of \citet{Mahdi2024}. Even when individual flux tubes cannot be spatially resolved, their formation and orbital motion induce measurable signatures, including an increase in the apparent size of the central emission region and periodic motion of the flux centroid. In Fig.~\ref{fig:unresolved_signatures}, we present these unresolved observational signatures associated with flux-tube formation and orbital motion, which may serve as templates for the analysis of future space-based VLBI observations.

\section{Discussion}
\label{sec:discussion}

We performed a Bayesian analysis of the EHT observations of Cen\,A directly in Fourier space, using synthetic visibilities generated from GRMHD simulations coupled with GRRT calculations. The radiative modeling self-consistently includes turbulent particle heating, radiative cooling, and anisotropic particle distribution functions. Overall, our models reproduce the EHT observables well, showing good agreement with both the visibility amplitudes and closure phases.

We find that all tested models exhibit residual discrepancies in the visibility amplitudes at baseline lengths of $\sim3.5\,\mathrm{G}\lambda$. These data points predominantly involve baselines including the Large Millimeter Telescope (LMT), which is known to have suffered from pointing issues during the Cen\,A observing campaign, leading to significant amplitude variability \citep{Janssen2021}. Furthermore, due to the substantial numerical cost of the Bayesian framework, we did not perform a dedicated gain calibration when comparing synthetic and observed visibilities. Incorporating gain calibration is expected to improve the quality of the fits and will be addressed in future, more advanced analyses.

Beyond the EHT total-intensity measurements, we further validated our model using independent observational constraints. In particular, we compared our simulations with polarimetric measurements from ALMA-only observations of Cen\,A, as well as with the spectral index inferred between 213\,GHz and 229\,GHz. Despite the smaller field of view (FOV) of our simulations, the model reproduces these observables remarkably well, lending strong support to the adopted radiation model and the underlying assumptions about plasma microphysics.

A detailed analysis of the jet collimation profile and limb stability reveals overall consistency with observational constraints. The inferred collimation profile agrees well with the measured jet shape, although the expansion exponents derived from our simulations, if considered without uncertainties, are systematically slightly larger than those inferred from observations. This discrepancy can be attributed to the initial conditions of our simulations, in which the ambient medium is represented by a numerical floor model and therefore does not exert significant external pressure capable of contributing to jet collimation. Future simulations incorporating a dense ambient medium are expected to provide additional collimation through external pressure confinement \citep{Rohoza2024}.

We also find that the jet base in our simulations is more compact than that inferred from reconstructed EHT images. This difference may arise from two distinct effects. First, the image reconstruction is based on the sparse $u$--$v$ coverage of the current EHT array, allowing multiple image morphologies to fit the data comparably well. Future EHT observations with an expanded array are expected to significantly reduce these degeneracies and yield more robust constraints on the jet base morphology. Second, our simulations assume a torus seeded with a weak, large-scale poloidal magnetic field, a configuration that favors jet launching via the BZ mechanism while producing only a weak disk wind. Alternative magnetic field geometries \citep[see, e.g.,][]{Dihingia2024} can simultaneously power BZ jets and stronger disk-driven outflows, such as BP winds. Exploring such configurations will be an important direction for future modeling efforts of Cen\,A.

Our analysis reveals a viewing angle of $\vartheta \sim 72^\circ$, which is significantly larger than estimates from lower-frequency and larger-FOV observations,
$12^\circ \leq \vartheta \leq 45^\circ$ \citep{Mueller2014}, however, is in agreement with large-scale estimates $50^\circ \leq \theta \leq 80^\circ$ \cite{Tingay_1998}. In our modeling, this larger viewing angle is primarily driven by the non-negligible
flux contribution of the counter-jet on scales up to $\sim 100\,\mu\mathrm{as}$. Adopting a smaller viewing angle would result in stronger Doppler
de-boosting of the counter-jet emission, which would require larger mass units to reproduce the observed flux density. However, increasing the mass
unit leads to higher optical depths in horizon-scale regions and enhances radiative losses (see the detailed discussion in
Sect.~\ref{sec: results MCMC}). This discrepancy between the inferred viewing angles, therefore, suggests a possible bend in the jet on larger scales,
which could be caused by (i) interaction with the ambient medium or (ii) the growth of instabilities during jet propagation.

\section{Summary \& Conclusion}
\label{sec:summary}
We present a novel framework that combines state-of-the-art GRMHD and GRRT simulations with high-resolution Event Horizon Telescope (EHT) observations to probe predictions from kinetic PIC simulations, specifically, the pitch-angle anisotropy of accelerated electron distribution functions. Our results demonstrate that the observed edge-brightening in Cen\,A can be naturally reproduced by an anisotropic particle distribution. We further identify observable signatures of this model, including the stability of the jet collimation profile and jet limbs, as well as the imprint of orbiting flux tubes, a distinctive signature of the magnetically arrested disk (MAD) accretion scenario. Future work should incorporate a more detailed treatment of particle anisotropy, i.e.\ $\eta(\sigma,\beta,\gamma_e)$, alongside improved reconstruction of synthetic visibilities and additional observational constraints, such as polarization fractions and broadband spectra, which will further tighten constraints on the underlying theoretical models.

\bibliographystyle{aa_url.bst}
\bibliography{bibliography}

@ARTICLE{EHT_M87_P3,
       author = {{EHT Collaboration} and {Akiyama}, Kazunori and {Alberdi}, Antxon and {Alef}, Walter and {Asada}, Keiichi and {Azulay}, Rebecca and {Baczko}, Anne-Kathrin and {Ball}, David and {Balokovi{\'c}}, Mislav and {Barrett}, John and {Bintley}, Dan and {Blackburn}, Lindy and {Boland}, Wilfred and {Bouman}, Katherine L. and {Bower}, Geoffrey C. and {Bremer}, Michael and {Brinkerink}, Christiaan D. and {Brissenden}, Roger and {Britzen}, Silke and {Broderick}, Avery E. and {Broguiere}, Dominique and {Bronzwaer}, Thomas and {Byun}, Do-Young and {Carlstrom}, John E. and {Chael}, Andrew and {Chan}, Chi-kwan and {Chatterjee}, Shami and {Chatterjee}, Koushik and {Chen}, Ming-Tang and {Chen}, Yongjun and {Cho}, Ilje and {Christian}, Pierre and {Conway}, John E. and {Cordes}, James M. and {Crew}, Geoffrey B. and {Cui}, Yuzhu and {Davelaar}, Jordy and {De Laurentis}, Mariafelicia and {Deane}, Roger and {Dempsey}, Jessica and {Desvignes}, Gregory and {Dexter}, Jason and {Doeleman}, Sheperd S. and {Eatough}, Ralph P. and {Falcke}, Heino and {Fish}, Vincent L. and {Fomalont}, Ed and {Fraga-Encinas}, Raquel and {Friberg}, Per and {Fromm}, Christian M. and {G{\'o}mez}, Jos{\'e} L. and {Galison}, Peter and {Gammie}, Charles F. and {Garc{\'\i}a}, Roberto and {Gentaz}, Olivier and {Georgiev}, Boris and {Goddi}, Ciriaco and {Gold}, Roman and {Gu}, Minfeng and {Gurwell}, Mark and {Hada}, Kazuhiro and {Hecht}, Michael H. and {Hesper}, Ronald and {Ho}, Luis C. and {Ho}, Paul and {Honma}, Mareki and {Huang}, Chih-Wei L. and {Huang}, Lei and {Hughes}, David H. and {Ikeda}, Shiro and {Inoue}, Makoto and {Issaoun}, Sara and {James}, David J. and {Jannuzi}, Buell T. and {Janssen}, Michael and {Jeter}, Britton and {Jiang}, Wu and {Johnson}, Michael D. and {Jorstad}, Svetlana and {Jung}, Taehyun and {Karami}, Mansour and {Karuppusamy}, Ramesh and {Kawashima}, Tomohisa and {Keating}, Garrett K. and {Kettenis}, Mark and {Kim}, Jae-Young and {Kim}, Junhan and {Kim}, Jongsoo and {Kino}, Motoki and {Koay}, Jun Yi and {Koch}, Patrick M. and {Koyama}, Shoko and {Kramer}, Michael and {Kramer}, Carsten and {Krichbaum}, Thomas P. and {Kuo}, Cheng-Yu and {Lauer}, Tod R. and {Lee}, Sang-Sung and {Li}, Yan-Rong and {Li}, Zhiyuan and {Lindqvist}, Michael and {Liu}, Kuo and {Liuzzo}, Elisabetta and {Lo}, Wen-Ping and {Lobanov}, Andrei P. and {Loinard}, Laurent and {Lonsdale}, Colin and {Lu}, Ru-Sen and {MacDonald}, Nicholas R. and {Mao}, Jirong and {Markoff}, Sera and {Marrone}, Daniel P. and {Marscher}, Alan P. and {Mart{\'\i}-Vidal}, Iv{\'a}n and {Matsushita}, Satoki and {Matthews}, Lynn D. and {Medeiros}, Lia and {Menten}, Karl M. and {Mizuno}, Yosuke and {Mizuno}, Izumi and {Moran}, James M. and {Moriyama}, Kotaro and {Moscibrodzka}, Monika and {M{\"u}ller}, Cornelia and {Nagai}, Hiroshi and {Nagar}, Neil M. and {Nakamura}, Masanori and {Narayan}, Ramesh and {Narayanan}, Gopal and {Natarajan}, Iniyan and {Neri}, Roberto and {Ni}, Chunchong and {Noutsos}, Aristeidis and {Okino}, Hiroki and {Olivares}, H{\'e}ctor and {Ortiz-Le{\'o}n}, Gisela N. and {Oyama}, Tomoaki and {{\"O}zel}, Feryal and {Palumbo}, Daniel C.~M. and {Patel}, Nimesh and {Pen}, Ue-Li and {Pesce}, Dominic W. and {Pi{\'e}tu}, Vincent and {Plambeck}, Richard and {PopStefanija}, Aleksandar and {Porth}, Oliver and {Prather}, Ben and {Preciado-L{\'o}pez}, Jorge A. and {Psaltis}, Dimitrios and {Pu}, Hung-Yi and {Ramakrishnan}, Venkatessh and {Rao}, Ramprasad and {Rawlings}, Mark G. and {Raymond}, Alexander W. and {Rezzolla}, Luciano and {Ripperda}, Bart and {Roelofs}, Freek and {Rogers}, Alan and {Ros}, Eduardo and {Rose}, Mel and {Roshanineshat}, Arash and {Rottmann}, Helge and {Roy}, Alan L. and {Ruszczyk}, Chet and {Ryan}, Benjamin R. and {Rygl}, Kazi L.~J. and {S{\'a}nchez}, Salvador and {S{\'a}nchez-Arguelles}, David and {Sasada}, Mahito and {Savolainen}, Tuomas and {Schloerb}, F. Peter and {Schuster}, Karl-Friedrich and {Shao}, Lijing and {Shen}, Zhiqiang and {Small}, Des and {Sohn}, Bong Won and {SooHoo}, Jason and {Tazaki}, Fumie and {Tiede}, Paul and {Tilanus}, Remo P.~J. and {Titus}, Michael and {Toma}, Kenji and {Torne}, Pablo and {Trent}, Tyler and {Trippe}, Sascha and {Tsuda}, Shuichiro and {van Bemmel}, Ilse and {van Langevelde}, Huib Jan and {van Rossum}, Daniel R. and {Wagner}, Jan and {Wardle}, John and {Weintroub}, Jonathan and {Wex}, Norbert and {Wharton}, Robert and {Wielgus}, Maciek and {Wong}, George N. and {Wu}, Qingwen and {Young}, Andr{\'e} and {Young}, Ken and {Younsi}, Ziri},
        title = "{First M87 Event Horizon Telescope Results. III. Data Processing and Calibration}",
      journal = {\apjl},
         year = 2019,
        month = apr,
       volume = {875},
       number = {1},
          eid = {L3},
        pages = {L3},
          doi = {10.3847/2041-8213/ab0c57},
archivePrefix = {arXiv},
       eprint = {1906.11240},
 primaryClass = {astro-ph.GA},
       adsurl = {https://ui.adsabs.harvard.edu/abs/2019ApJ...875L...3E}
}

@ARTICLE{EHT_M87_P5,
       author = {{EHT Collaboration} and {Akiyama}, Kazunori and {Alberdi}, Antxon and {Alef}, Walter and {Asada}, Keiichi and {Azulay}, Rebecca and {Baczko}, Anne-Kathrin and {Ball}, David and {Balokovi{\'c}}, Mislav and {Barrett}, John and {Bintley}, Dan and {Blackburn}, Lindy and {Boland}, Wilfred and {Bouman}, Katherine L. and {Bower}, Geoffrey C. and {Bremer}, Michael and {Brinkerink}, Christiaan D. and {Brissenden}, Roger and {Britzen}, Silke and {Broderick}, Avery E. and {Broguiere}, Dominique and {Bronzwaer}, Thomas and {Byun}, Do-Young and {Carlstrom}, John E. and {Chael}, Andrew and {Chan}, Chi-kwan and {Chatterjee}, Shami and {Chatterjee}, Koushik and {Chen}, Ming-Tang and {Chen}, Yongjun and {Cho}, Ilje and {Christian}, Pierre and {Conway}, John E. and {Cordes}, James M. and {Crew}, Geoffrey B. and {Cui}, Yuzhu and {Davelaar}, Jordy and {De Laurentis}, Mariafelicia and {Deane}, Roger and {Dempsey}, Jessica and {Desvignes}, Gregory and {Dexter}, Jason and {Doeleman}, Sheperd S. and {Eatough}, Ralph P. and {Falcke}, Heino and {Fish}, Vincent L. and {Fomalont}, Ed and {Fraga-Encinas}, Raquel and {Friberg}, Per and {Fromm}, Christian M. and {G{\'o}mez}, Jos{\'e} L. and {Galison}, Peter and {Gammie}, Charles F. and {Garc{\'\i}a}, Roberto and {Gentaz}, Olivier and {Georgiev}, Boris and {Goddi}, Ciriaco and {Gold}, Roman and {Gu}, Minfeng and {Gurwell}, Mark and {Hada}, Kazuhiro and {Hecht}, Michael H. and {Hesper}, Ronald and {Ho}, Luis C. and {Ho}, Paul and {Honma}, Mareki and {Huang}, Chih-Wei L. and {Huang}, Lei and {Hughes}, David H. and {Ikeda}, Shiro and {Inoue}, Makoto and {Issaoun}, Sara and {James}, David J. and {Jannuzi}, Buell T. and {Janssen}, Michael and {Jeter}, Britton and {Jiang}, Wu and {Johnson}, Michael D. and {Jorstad}, Svetlana and {Jung}, Taehyun and {Karami}, Mansour and {Karuppusamy}, Ramesh and {Kawashima}, Tomohisa and {Keating}, Garrett K. and {Kettenis}, Mark and {Kim}, Jae-Young and {Kim}, Junhan and {Kim}, Jongsoo and {Kino}, Motoki and {Koay}, Jun Yi and {Koch}, Patrick M. and {Koyama}, Shoko and {Kramer}, Michael and {Kramer}, Carsten and {Krichbaum}, Thomas P. and {Kuo}, Cheng-Yu and {Lauer}, Tod R. and {Lee}, Sang-Sung and {Li}, Yan-Rong and {Li}, Zhiyuan and {Lindqvist}, Michael and {Liu}, Kuo and {Liuzzo}, Elisabetta and {Lo}, Wen-Ping and {Lobanov}, Andrei P. and {Loinard}, Laurent and {Lonsdale}, Colin and {Lu}, Ru-Sen and {MacDonald}, Nicholas R. and {Mao}, Jirong and {Markoff}, Sera and {Marrone}, Daniel P. and {Marscher}, Alan P. and {Mart{\'\i}-Vidal}, Iv{\'a}n and {Matsushita}, Satoki and {Matthews}, Lynn D. and {Medeiros}, Lia and {Menten}, Karl M. and {Mizuno}, Yosuke and {Mizuno}, Izumi and {Moran}, James M. and {Moriyama}, Kotaro and {Moscibrodzka}, Monika and {Mul{\ensuremath{\ddot{}}}ler}, Cornelia and {Nagai}, Hiroshi and {Nagar}, Neil M. and {Nakamura}, Masanori and {Narayan}, Ramesh and {Narayanan}, Gopal and {Natarajan}, Iniyan and {Neri}, Roberto and {Ni}, Chunchong and {Noutsos}, Aristeidis and {Okino}, Hiroki and {Olivares}, H{\'e}ctor and {Oyama}, Tomoaki and {{\"O}zel}, Feryal and {Palumbo}, Daniel C.~M. and {Patel}, Nimesh and {Pen}, Ue-Li and {Pesce}, Dominic W. and {Pi{\'e}tu}, Vincent and {Plambeck}, Richard and {PopStefanija}, Aleksandar and {Porth}, Oliver and {Prather}, Ben and {Preciado-L{\'o}pez}, Jorge A. and {Psaltis}, Dimitrios and {Pu}, Hung-Yi and {Ramakrishnan}, Venkatessh and {Rao}, Ramprasad and {Rawlings}, Mark G. and {Raymond}, Alexander W. and {Rezzolla}, Luciano and {Ripperda}, Bart and {Roelofs}, Freek and {Rogers}, Alan and {Ros}, Eduardo and {Rose}, Mel and {Roshanineshat}, Arash and {Rottmann}, Helge and {Roy}, Alan L. and {Ruszczyk}, Chet and {Ryan}, Benjamin R. and {Rygl}, Kazi L.~J. and {S{\'a}nchez}, Salvador and {S{\'a}nchez-Arguelles}, David and {Sasada}, Mahito and {Savolainen}, Tuomas and {Schloerb}, F. Peter and {Schuster}, Karl-Friedrich and {Shao}, Lijing and {Shen}, Zhiqiang and {Small}, Des and {Sohn}, Bong Won and {SooHoo}, Jason and {Tazaki}, Fumie and {Tiede}, Paul and {Tilanus}, Remo P.~J. and {Titus}, Michael and {Toma}, Kenji and {Torne}, Pablo and {Trent}, Tyler and {Trippe}, Sascha and {Tsuda}, Shuichiro and {van Bemmel}, Ilse and {van Langevelde}, Huib Jan and {van Rossum}, Daniel R. and {Wagner}, Jan and {Wardle}, John and {Weintroub}, Jonathan and {Wex}, Norbert and {Wharton}, Robert and {Wielgus}, Maciek and {Wong}, George N. and {Wu}, Qingwen and {Young}, Andr{\'e} and {Young}, Ken and {Younsi}, Ziri and {Yuan}, Feng},
        title = "{First M87 Event Horizon Telescope Results. V. Physical Origin of the Asymmetric Ring}",
      journal = {\apjl},
         year = 2019,
        month = apr,
       volume = {875},
       number = {1},
          eid = {L5},
        pages = {L5},
          doi = {10.3847/2041-8213/ab0f43},
archivePrefix = {arXiv},
       eprint = {1906.11242},
 primaryClass = {astro-ph.GA},
       adsurl = {https://ui.adsabs.harvard.edu/abs/2019ApJ...875L...5E}
}

@ARTICLE{BZ1977,
       author = {{Blandford}, R.~D. and {Znajek}, R.~L.},
        title = "{Electromagnetic extraction of energy from Kerr black holes.}",
      journal = {\mnras},
         year = 1977,
        month = may,
       volume = {179},
        pages = {433-456},
          doi = {10.1093/mnras/179.3.433},
       adsurl = {https://ui.adsabs.harvard.edu/abs/1977MNRAS.179..433B}
}

@ARTICLE{Scepi2022,
       author = {{Scepi}, Nicolas and {Dexter}, Jason and {Begelman}, Mitchell C.},
        title = "{Sgr A* X-ray flares from non-thermal particle acceleration in a magnetically arrested disc}",
      journal = {\mnras},
         year = 2022,
        month = apr,
       volume = {511},
       number = {3},
        pages = {3536-3547},
          doi = {10.1093/mnras/stac337},
archivePrefix = {arXiv},
       eprint = {2107.08056},
 primaryClass = {astro-ph.HE},
       adsurl = {https://ui.adsabs.harvard.edu/abs/2022MNRAS.511.3536S}
}

@ARTICLE{Hu2025,
       author = {{Hu}, Xu-Fan and {Jiang}, Hong-Xuan and {Mizuno}, Yosuke and {Fromm}, Christian M. and {Vaidya}, Bhargav},
        title = "{Quasiperiodic Polarized Emissions from Kink Structure in Magnetized Relativistic Jets}",
      journal = {\apj},
         year = 2025,
        month = dec,
       volume = {995},
       number = {1},
          eid = {76},
        pages = {76},
          doi = {10.3847/1538-4357/ae1a6f},
archivePrefix = {arXiv},
       eprint = {2511.03140},
 primaryClass = {astro-ph.HE},
       adsurl = {https://ui.adsabs.harvard.edu/abs/2025ApJ...995...76H}
}

@ARTICLE{BP1982,
       author = {{Blandford}, R.~D. and {Payne}, D.~G.},
        title = "{Hydromagnetic flows from accretion disks and the production of radio jets.}",
      journal = {\mnras},
         year = 1982,
        month = jun,
       volume = {199},
        pages = {883-903},
          doi = {10.1093/mnras/199.4.883},
       adsurl = {https://ui.adsabs.harvard.edu/abs/1982MNRAS.199..883B}
}

@ARTICLE{Mizuno2007,
       author = {{Mizuno}, Yosuke and {Hardee}, Philip and {Nishikawa}, Ken-Ichi},
        title = "{Three-dimensional Relativistic Magnetohydrodynamic Simulations of Magnetized Spine-Sheath Relativistic Jets}",
      journal = {\apj},
         year = 2007,
        month = jun,
       volume = {662},
       number = {2},
        pages = {835-850},
          doi = {10.1086/518106},
archivePrefix = {arXiv},
       eprint = {astro-ph/0703190},
 primaryClass = {astro-ph},
       adsurl = {https://ui.adsabs.harvard.edu/abs/2007ApJ...662..835M}
}

@INPROCEEDINGS{Johnson2024,
       author = {{Johnson}, Michael D. and {Akiyama}, Kazunori and {Baturin}, Rebecca and {Bilyeu}, Bryan and {Blackburn}, Lindy and {Boroson}, Don and {C{\'a}rdenas-Avenda{\~n}o}, Alejandro and {Chael}, Andrew and {Chan}, Chi-kwan and {Chang}, Dominic and {Cheimets}, Peter and {Chou}, Cathy and {Doeleman}, Sheperd S. and {Farah}, Joseph and {Galison}, Peter and {Gamble}, Ronald and {Gammie}, Charles F. and {Gelles}, Zachary and {G{\'o}mez}, Jos{\'e} L. and {Gralla}, Samuel E. and {Grimes}, Paul and {Gurvits}, Leonid I. and {Hadar}, Shahar and {Haworth}, Kari and {Hada}, Kazuhiro and {Hecht}, Michael H. and {Honma}, Mareki and {Houston}, Janice and {Hudson}, Ben and {Issaoun}, Sara and {Jia}, He and {Jorstad}, Svetlana and {Kauffman}, Jens and {Kovalev}, Yuri Y. and {Kurczynski}, Peter and {Lafon}, Robert E. and {Lupsasca}, Alexandru and {Lehmensiek}, Robert and {Ma}, Chung-Pei and {Marrone}, Daniel P. and {Marscher}, Alan P. and {Melnick}, Gary and {Narayan}, Ramesh and {Niinuma}, Kotaro and {Noble}, Scott C. and {Palmer}, Eric J. and {Palumbo}, Daniel C.~M. and {Paritsky}, Lenny and {Peretz}, Eliad and {Pesce}, Dominic and {Plavin}, Alexander and {Quataert}, Eliot and {Rana}, Hannah and {Ricarte}, Angelo and {Roelofs}, Freek and {Shtyrkova}, Katia and {Sinclair}, Laura C. and {Small}, Jeffrey and {Kumara}, Sridharan Tirupati and {Srinivasan}, Ranjani and {Strominger}, Andrew and {Tiede}, Paul and {Tong}, Edward and {Wang}, Jade and {Weintroub}, Jonathan and {Wielgus}, Maciek and {Wong}, George},
        title = "{The Black Hole Explorer: motivation and vision}",
    booktitle = {Space Telescopes and Instrumentation 2024: Optical, Infrared, and Millimeter Wave},
         year = 2024,
       editor = {{Coyle}, Laura E. and {Matsuura}, Shuji and {Perrin}, Marshall D.},
       series = {Society of Photo-Optical Instrumentation Engineers (SPIE) Conference Series},
       volume = {13092},
        month = aug,
          eid = {130922D},
        pages = {130922D},
          doi = {10.1117/12.3019835},
archivePrefix = {arXiv},
       eprint = {2406.12917},
 primaryClass = {astro-ph.IM},
       adsurl = {https://ui.adsabs.harvard.edu/abs/2024SPIE13092E..2DJ}
}

@ARTICLE{Dihingia2024,
       author = {{Dihingia}, Indu K. and {Fendt}, Christian},
        title = "{Thin Accretion disks in GR-MHD simulations}",
      journal = {arXiv e-prints},
         year = 2024,
        month = apr,
          eid = {arXiv:2404.06140},
        pages = {arXiv:2404.06140},
          doi = {10.48550/arXiv.2404.06140},
archivePrefix = {arXiv},
       eprint = {2404.06140},
 primaryClass = {astro-ph.HE},
       adsurl = {https://ui.adsabs.harvard.edu/abs/2024arXiv240406140D}
}

@ARTICLE{Rohoza2024,
       author = {{Rohoza}, Valeriia and {Lalakos}, Aretaios and {Paik}, Max and {Chatterjee}, Koushik and {Liska}, Matthew and {Tchekhovskoy}, Alexander and {Gottlieb}, Ore},
        title = "{How to Turn Jets into Cylinders near Supermassive Black Holes in 3D General Relativistic Magnetohydrodynamic Simulations}",
      journal = {\apjl},
         year = 2024,
        month = mar,
       volume = {963},
       number = {1},
          eid = {L29},
        pages = {L29},
          doi = {10.3847/2041-8213/ad24fc},
       adsurl = {https://ui.adsabs.harvard.edu/abs/2024ApJ...963L..29R}
}

@ARTICLE{Janssen2021,
       author = {{Janssen}, Michael and {Falcke}, Heino and {Kadler}, Matthias and {Ros}, Eduardo and {Wielgus}, Maciek and {Akiyama}, Kazunori and {Balokovi{\'c}}, Mislav and {Blackburn}, Lindy and {Bouman}, Katherine L. and {Chael}, Andrew and {Chan}, Chi-kwan and {Chatterjee}, Koushik and {Davelaar}, Jordy and {Edwards}, Philip G. and {Fromm}, Christian M. and {G{\'o}mez}, Jos{\'e} L. and {Goddi}, Ciriaco and {Issaoun}, Sara and {Johnson}, Michael D. and {Kim}, Junhan and {Koay}, Jun Yi and {Krichbaum}, Thomas P. and {Liu}, Jun and {Liuzzo}, Elisabetta and {Markoff}, Sera and {Markowitz}, Alex and {Marrone}, Daniel P. and {Mizuno}, Yosuke and {M{\"u}ller}, Cornelia and {Ni}, Chunchong and {Pesce}, Dominic W. and {Ramakrishnan}, Venkatessh and {Roelofs}, Freek and {Rygl}, Kazi L.~J. and {van Bemmel}, Ilse and {Event Horizon Telescope Collaboration} and {Alberdi}, Antxon and {Alef}, Walter and {Algaba}, Juan Carlos and {Anantua}, Richard and {Asada}, Keiichi and {Azulay}, Rebecca and {Baczko}, Anne-Kathrin and {Ball}, David and {Ball}, David and {Barrett}, John and {Benson}, Bradford A. and {Bintley}, Dan and {Bintley}, Dan and {Blundell}, Raymond and {Boland}, Wilfred and {Boland}, Wilfred and {Bower}, Geoffrey C. and {Boyce}, Hope and {Bremer}, Michael and {Brinkerink}, Christiaan D. and {Brissenden}, Roger and {Britzen}, Silke and {Broderick}, Avery E. and {Broguiere}, Dominique and {Bronzwaer}, Thomas and {Byun}, Do-Young and {Carlstrom}, John E. and {Chatterjee}, Shami and {Chen}, Ming-Tang and {Chen}, Yongjun and {Chesler}, Paul M. and {Cho}, Ilje and {Christian}, Pierre and {Conway}, John E. and {Cordes}, James M. and {Crawford}, Thomas M. and {Crew}, Geoffrey B. and {Cruz-Osorio}, Alejandro and {Cui}, Yuzhu and {Cui}, Yuzhu and {De Laurentis}, Mariafelicia and {Deane}, Roger and {Dempsey}, Jessica and {Desvignes}, Gregory and {Dexter}, Jason and {Doeleman}, Sheperd S. and {Eatough}, Ralph P. and {Farah}, Joseph and {Farah}, Joseph and {Fish}, Vincent L. and {Fomalont}, Ed and {Ford}, H. Alyson and {Fraga-Encinas}, Raquel and {Friberg}, Per and {Friberg}, Per and {Fuentes}, Antonio and {Galison}, Peter and {Gammie}, Charles F. and {Garc{\'\i}a}, Roberto and {Gelles}, Zachary and {Gentaz}, Olivier and {Georgiev}, Boris and {Georgiev}, Boris and {Gold}, Roman and {Gold}, Roman and {G{\'o}mez-Ruiz}, Arturo I. and {Gu}, Minfeng and {Gurwell}, Mark and {Hada}, Kazuhiro and {Haggard}, Daryl and {Hecht}, Michael H. and {Hesper}, Ronald and {Himwich}, Elizabeth and {Ho}, Luis C. and {Ho}, Paul and {Honma}, Mareki and {Huang}, Chih-Wei L. and {Huang}, Lei and {Hughes}, David H. and {Ikeda}, Shiro and {Inoue}, Makoto and {Inoue}, Makoto and {James}, David J. and {Jannuzi}, Buell T. and {Jeter}, Britton and {Jiang}, Wu and {Jimenez-Rosales}, Alejandra and {Jorstad}, Svetlana and {Jung}, Taehyun and {Karami}, Mansour and {Karuppusamy}, Ramesh and {Kawashima}, Tomohisa and {Keating}, Garrett K. and {Kettenis}, Mark and {Kim}, Dong-Jin and {Kim}, Jae-Young and {Kim}, Jongsoo and {Kino}, Motoki and {Kofuji}, Yutaro and {Koyama}, Shoko and {Kramer}, Michael and {Kramer}, Carsten and {Kuo}, Cheng-Yu and {Lauer}, Tod R. and {Lee}, Sang-Sung and {Levis}, Aviad and {Li}, Yan-Rong and {Li}, Zhiyuan and {Lindqvist}, Michael and {Lico}, Rocco and {Lindahl}, Greg and {Liu}, Kuo and {Lo}, Wen-Ping and {Lobanov}, Andrei P. and {Loinard}, Laurent and {Lonsdale}, Colin and {Lu}, Ru-Sen and {MacDonald}, Nicholas R. and {Mao}, Jirong and {Marchili}, Nicola and {Marscher}, Alan P. and {Mart{\'\i}-Vidal}, Iv{\'a}n and {Matsushita}, Satoki and {Matthews}, Lynn D. and {Medeiros}, Lia and {Menten}, Karl M. and {Mizuno}, Izumi and {Moran}, James M. and {Moriyama}, Kotaro and {Moscibrodzka}, Monika and {Moscibrodzka}, Monika and {Musoke}, Gibwa and {Mej{\'\i}as}, Alejandro Mus and {Nagai}, Hiroshi and {Nagar}, Neil M. and {Nakamura}, Masanori and {Narayan}, Ramesh and {Narayanan}, Gopal and {Natarajan}, Iniyan and {Nathanail}, Antonios and {Neilsen}, Joey and {Neri}, Roberto and {Noutsos}, Aristeidis and {Nowak}, Michael A. and {Okino}, Hiroki and {Olivares}, H{\'e}ctor and {Ortiz-Le{\'o}n}, Gisela N. and {Oyama}, Tomoaki and {{\"O}zel}, Feryal and {Palumbo}, Daniel C.~M. and {Park}, Jongho and {Patel}, Nimesh and {Pen}, Ue-Li and {Pi{\'e}tu}, Vincent and {Plambeck}, Richard and {PopStefanija}, Aleksandar and {Porth}, Oliver and {P{\"o}tzl}, Felix M. and {Prather}, Ben and {Preciado-L{\'o}pez}, Jorge A. and {Psaltis}, Dimitrios and {Pu}, Hung-Yi and {Pu}, Hung-Yi and {Rao}, Ramprasad},
        title = "{Event Horizon Telescope observations of the jet launching and collimation in Centaurus A}",
      journal = {Nature Astronomy},
         year = 2021,
        month = jul,
       volume = {5},
        pages = {1017-1028},
          doi = {10.1038/s41550-021-01417-w},
archivePrefix = {arXiv},
       eprint = {2111.03356},
 primaryClass = {astro-ph.GA},
       adsurl = {https://ui.adsabs.harvard.edu/abs/2021NatAs...5.1017J}
}

@ARTICLE{Mueller2014,
       author = {{M{\"u}ller}, C. and {Kadler}, M. and {Ojha}, R. and {Perucho}, M. and {Gro{\ss}berger}, C. and {Ros}, E. and {Wilms}, J. and {Blanchard}, J. and {B{\"o}ck}, M. and {Carpenter}, B. and {Dutka}, M. and {Edwards}, P.~G. and {Hase}, H. and {Horiuchi}, S. and {Kreikenbohm}, A. and {Lovell}, J.~E.~J. and {Markowitz}, A. and {Phillips}, C. and {Pl{\"o}tz}, C. and {Pursimo}, T. and {Quick}, J. and {Rothschild}, R. and {Schulz}, R. and {Steinbring}, T. and {Stevens}, J. and {Tr{\"u}stedt}, J. and {Tzioumis}, A.~K.},
        title = "{TANAMI monitoring of Centaurus A: The complex dynamics in the inner parsec of an extragalactic jet}",
      journal = {\aap},
         year = 2014,
        month = sep,
       volume = {569},
          eid = {A115},
        pages = {A115},
          doi = {10.1051/0004-6361/201423948},
archivePrefix = {arXiv},
       eprint = {1407.0162},
 primaryClass = {astro-ph.HE},
       adsurl = {https://ui.adsabs.harvard.edu/abs/2014A&A...569A.115M}
}

@ARTICLE{EHT_M87_P1,
       author = {{EHT Collaboration} and {Akiyama}, Kazunori and {Alberdi}, Antxon and {Alef}, Walter and {Asada}, Keiichi and {Azulay}, Rebecca and {Baczko}, Anne-Kathrin and {Ball}, David and {Balokovi{\'c}}, Mislav and {Barrett}, John and {Bintley}, Dan and {Blackburn}, Lindy and {Boland}, Wilfred and {Bouman}, Katherine L. and {Bower}, Geoffrey C. and {Bremer}, Michael and {Brinkerink}, Christiaan D. and {Brissenden}, Roger and {Britzen}, Silke and {Broderick}, Avery E. and {Broguiere}, Dominique and {Bronzwaer}, Thomas and {Byun}, Do-Young and {Carlstrom}, John E. and {Chael}, Andrew and {Chan}, Chi-kwan and {Chatterjee}, Shami and {Chatterjee}, Koushik and {Chen}, Ming-Tang and {Chen}, Yongjun and {Cho}, Ilje and {Christian}, Pierre and {Conway}, John E. and {Cordes}, James M. and {Crew}, Geoffrey B. and {Cui}, Yuzhu and {Davelaar}, Jordy and {De Laurentis}, Mariafelicia and {Deane}, Roger and {Dempsey}, Jessica and {Desvignes}, Gregory and {Dexter}, Jason and {Doeleman}, Sheperd S. and {Eatough}, Ralph P. and {Falcke}, Heino and {Fish}, Vincent L. and {Fomalont}, Ed and {Fraga-Encinas}, Raquel and {Freeman}, William T. and {Friberg}, Per and {Fromm}, Christian M. and {G{\'o}mez}, Jos{\'e} L. and {Galison}, Peter and {Gammie}, Charles F. and {Garc{\'\i}a}, Roberto and {Gentaz}, Olivier and {Georgiev}, Boris and {Goddi}, Ciriaco and {Gold}, Roman and {Gu}, Minfeng and {Gurwell}, Mark and {Hada}, Kazuhiro and {Hecht}, Michael H. and {Hesper}, Ronald and {Ho}, Luis C. and {Ho}, Paul and {Honma}, Mareki and {Huang}, Chih-Wei L. and {Huang}, Lei and {Hughes}, David H. and {Ikeda}, Shiro and {Inoue}, Makoto and {Issaoun}, Sara and {James}, David J. and {Jannuzi}, Buell T. and {Janssen}, Michael and {Jeter}, Britton and {Jiang}, Wu and {Johnson}, Michael D. and {Jorstad}, Svetlana and {Jung}, Taehyun and {Karami}, Mansour and {Karuppusamy}, Ramesh and {Kawashima}, Tomohisa and {Keating}, Garrett K. and {Kettenis}, Mark and {Kim}, Jae-Young and {Kim}, Junhan and {Kim}, Jongsoo and {Kino}, Motoki and {Koay}, Jun Yi and {Koch}, Patrick M. and {Koyama}, Shoko and {Kramer}, Michael and {Kramer}, Carsten and {Krichbaum}, Thomas P. and {Kuo}, Cheng-Yu and {Lauer}, Tod R. and {Lee}, Sang-Sung and {Li}, Yan-Rong and {Li}, Zhiyuan and {Lindqvist}, Michael and {Liu}, Kuo and {Liuzzo}, Elisabetta and {Lo}, Wen-Ping and {Lobanov}, Andrei P. and {Loinard}, Laurent and {Lonsdale}, Colin and {Lu}, Ru-Sen and {MacDonald}, Nicholas R. and {Mao}, Jirong and {Markoff}, Sera and {Marrone}, Daniel P. and {Marscher}, Alan P. and {Mart{\'\i}-Vidal}, Iv{\'a}n and {Matsushita}, Satoki and {Matthews}, Lynn D. and {Medeiros}, Lia and {Menten}, Karl M. and {Mizuno}, Yosuke and {Mizuno}, Izumi and {Moran}, James M. and {Moriyama}, Kotaro and {Moscibrodzka}, Monika and {M{\"u}ller}, Cornelia and {Nagai}, Hiroshi and {Nagar}, Neil M. and {Nakamura}, Masanori and {Narayan}, Ramesh and {Narayanan}, Gopal and {Natarajan}, Iniyan and {Neri}, Roberto and {Ni}, Chunchong and {Noutsos}, Aristeidis and {Okino}, Hiroki and {Olivares}, H{\'e}ctor and {Ortiz-Le{\'o}n}, Gisela N. and {Oyama}, Tomoaki and {{\"O}zel}, Feryal and {Palumbo}, Daniel C.~M. and {Patel}, Nimesh and {Pen}, Ue-Li and {Pesce}, Dominic W. and {Pi{\'e}tu}, Vincent and {Plambeck}, Richard and {PopStefanija}, Aleksandar and {Porth}, Oliver and {Prather}, Ben and {Preciado-L{\'o}pez}, Jorge A. and {Psaltis}, Dimitrios and {Pu}, Hung-Yi and {Ramakrishnan}, Venkatessh and {Rao}, Ramprasad and {Rawlings}, Mark G. and {Raymond}, Alexander W. and {Rezzolla}, Luciano and {Ripperda}, Bart and {Roelofs}, Freek and {Rogers}, Alan and {Ros}, Eduardo and {Rose}, Mel and {Roshanineshat}, Arash and {Rottmann}, Helge and {Roy}, Alan L. and {Ruszczyk}, Chet and {Ryan}, Benjamin R. and {Rygl}, Kazi L.~J. and {S{\'a}nchez}, Salvador and {S{\'a}nchez-Arguelles}, David and {Sasada}, Mahito and {Savolainen}, Tuomas and {Schloerb}, F. Peter and {Schuster}, Karl-Friedrich and {Shao}, Lijing and {Shen}, Zhiqiang and {Small}, Des and {Sohn}, Bong Won and {SooHoo}, Jason and {Tazaki}, Fumie and {Tiede}, Paul and {Tilanus}, Remo P.~J. and {Titus}, Michael and {Toma}, Kenji and {Torne}, Pablo and {Trent}, Tyler and {Trippe}, Sascha and {Tsuda}, Shuichiro and {van Bemmel}, Ilse and {van Langevelde}, Huib Jan and {van Rossum}, Daniel R. and {Wagner}, Jan and {Wardle}, John and {Weintroub}, Jonathan and {Wex}, Norbert and {Wharton}, Robert and {Wielgus}, Maciek and {Wong}, George N. and {Wu}, Qingwen and {Young}, Ken and {Young}, Andr{\'e}},
        title = "{First M87 Event Horizon Telescope Results. I. The Shadow of the Supermassive Black Hole}",
      journal = {\apjl},
         year = 2019,
        month = apr,
       volume = {875},
       number = {1},
          eid = {L1},
        pages = {L1},
          doi = {10.3847/2041-8213/ab0ec7},
archivePrefix = {arXiv},
       eprint = {1906.11238},
 primaryClass = {astro-ph.GA},
       adsurl = {https://ui.adsabs.harvard.edu/abs/2019ApJ...875L...1E}
}

@BOOK{Meier2012,
       author = {{Meier}, David L.},
        title = "{Black Hole Astrophysics: The Engine Paradigm}",
         year = 2012,
          doi = {10.1007/978-3-642-01936-4},
       adsurl = {https://ui.adsabs.harvard.edu/abs/2012bhae.book.....M}
}

@ARTICLE{Rothschild2011,
       author = {{Rothschild}, R.~E. and {Markowitz}, A. and {Rivers}, E. and {Suchy}, S. and {Pottschmidt}, K. and {Kadler}, M. and {M{\"u}ller}, C. and {Wilms}, J.},
        title = "{Twelve and a Half Years of Observations of Centaurus a with the Rossi X-Ray Timing Explorer}",
      journal = {\apj},
         year = 2011,
        month = may,
       volume = {733},
       number = {1},
          eid = {23},
        pages = {23},
          doi = {10.1088/0004-637X/733/1/23},
archivePrefix = {arXiv},
       eprint = {1102.5076},
 primaryClass = {astro-ph.HE},
       adsurl = {https://ui.adsabs.harvard.edu/abs/2011ApJ...733...23R}
}

@ARTICLE{Fromm2022,
       author = {{Fromm}, Christian M. and {Cruz-Osorio}, Alejandro and {Mizuno}, Yosuke and {Nathanail}, Antonios and {Younsi}, Ziri and {Porth}, Oliver and {Olivares}, Hector and {Davelaar}, Jordy and {Falcke}, Heino and {Kramer}, Michael and {Rezzolla}, Luciano},
        title = "{Impact of non-thermal particles on the spectral and structural properties of M87}",
      journal = {\aap},
         year = 2022,
        month = apr,
       volume = {660},
          eid = {A107},
        pages = {A107},
          doi = {10.1051/0004-6361/202142295},
archivePrefix = {arXiv},
       eprint = {2111.02518},
 primaryClass = {astro-ph.HE},
       adsurl = {https://ui.adsabs.harvard.edu/abs/2022A&A...660A.107F}
}

@BOOK{Pacholczyk1970,
       author = {{Pacholczyk}, A.~G.},
        title = "{Radio astrophysics. Nonthermal processes in galactic and extragalactic sources}",
         year = 1970,
       adsurl = {https://ui.adsabs.harvard.edu/abs/1970ranp.book.....P}
}

@misc{prather2024,
  title         = {KHARMA: Flexible, Portable Performance for GRMHD},
  author        = {Ben S. Prather},
  year          = {2024},
  eprint        = {2408.01361},
  archiveprefix = {arXiv},
  primaryclass  = {astro-ph.HE},
  adsurl        = {https://doi.org/10.48550/arXiv.2408.01361}
}

@article{Prather2021,
  doi       = {10.21105/joss.03336},
  adsurl    = {https://doi.org/10.21105/joss.03336},
  year      = {2021},
  publisher = {The Open Journal},
  volume    = {6},
  number    = {66},
  pages     = {3336},
  author    = {Ben S. Prather and George N. Wong and Vedant Dhruv and Benjamin R. Ryan and Joshua C. Dolence and Sean M. Ressler and Charles F. Gammie},
  title     = {iharm3D: Vectorized General Relativistic Magnetohydrodynamics},
  journal   = {Journal of Open Source Software}
}

@ARTICLE{Comisso2024,
       author = {{Comisso}, Luca},
        title = "{Concurrent Particle Acceleration and Pitch-angle Anisotropy Driven by Magnetic Reconnection: Ion-electron Plasmas}",
      journal = {\apj},
         year = 2024,
        month = sep,
       volume = {972},
       number = {1},
          eid = {9},
        pages = {9},
          doi = {10.3847/1538-4357/ad51fe},
archivePrefix = {arXiv},
       eprint = {2405.18227},
 primaryClass = {astro-ph.HE},
       adsurl = {https://ui.adsabs.harvard.edu/abs/2024ApJ...972....9C}
}

@ARTICLE{FM1976,
	author = {{Fishbone}, L.~G. and {Moncrief}, V.},
	title = "{Relativistic fluid disks in orbit around Kerr black holes.}",
	journal = {\apj},
	year = 1976,
	month = aug,
	volume = {207},
	pages = {962-976},
	doi = {10.1086/154565},
	adsurl = {https://ui.adsabs.harvard.edu/abs/1976ApJ...207..962F}
}

@ARTICLE{Foreman2013,
       author = {{Foreman-Mackey}, Daniel and {Hogg}, David W. and {Lang}, Dustin and {Goodman}, Jonathan},
        title = "{emcee: The MCMC Hammer}",
      journal = {\pasp},
         year = 2013,
        month = mar,
       volume = {125},
       number = {925},
        pages = {306},
          doi = {10.1086/670067},
archivePrefix = {arXiv},
       eprint = {1202.3665},
 primaryClass = {astro-ph.IM},
       adsurl = {https://ui.adsabs.harvard.edu/abs/2013PASP..125..306F}
}

@ARTICLE{Davelaar2019,
       author = {{Davelaar}, Jordy and {Olivares}, Hector and {Porth}, Oliver and {Bronzwaer}, Thomas and {Janssen}, Michael and {Roelofs}, Freek and {Mizuno}, Yosuke and {Fromm}, Christian M. and {Falcke}, Heino and {Rezzolla}, Luciano},
        title = "{Modeling non-thermal emission from the jet-launching region of M 87 with adaptive mesh refinement}",
      journal = {\aap},
         year = 2019,
        month = dec,
       volume = {632},
          eid = {A2},
        pages = {A2},
          doi = {10.1051/0004-6361/201936150},
archivePrefix = {arXiv},
       eprint = {1906.10065},
 primaryClass = {astro-ph.HE},
       adsurl = {https://ui.adsabs.harvard.edu/abs/2019A&A...632A...2D}
}

@ARTICLE{Monika2018,
       author = {{Mo{\'s}cibrodzka}, M. and {Gammie}, C.~F.},
        title = "{IPOLE - semi-analytic scheme for relativistic polarized radiative transport}",
      journal = {\mnras},
         year = 2018,
        month = mar,
       volume = {475},
       number = {1},
        pages = {43-54},
          doi = {10.1093/mnras/stx3162},
archivePrefix = {arXiv},
       eprint = {1712.03057},
 primaryClass = {astro-ph.HE},
       adsurl = {https://ui.adsabs.harvard.edu/abs/2018MNRAS.475...43M}
}

@ARTICLE{Chatterjee2021,
       author = {{Chatterjee}, K. and {Markoff}, S. and {Neilsen}, J. and {Younsi}, Z. and {Witzel}, G. and {Tchekhovskoy}, A. and {Yoon}, D. and {Ingram}, A. and {van der Klis}, M. and {Boyce}, H. and {Do}, T. and {Haggard}, D. and {Nowak}, M.~A.},
        title = "{General relativistic MHD simulations of non-thermal flaring in Sagittarius A*}",
      journal = {\mnras},
         year = 2021,
        month = nov,
       volume = {507},
       number = {4},
        pages = {5281-5302},
          doi = {10.1093/mnras/stab2466},
archivePrefix = {arXiv},
       eprint = {2011.08904},
 primaryClass = {astro-ph.HE},
       adsurl = {https://ui.adsabs.harvard.edu/abs/2021MNRAS.507.5281C}
}

@ARTICLE{tchekhovskoy2011,
	author = {{Tchekhovskoy}, Alexander and {Narayan}, Ramesh and {McKinney}, Jonathan C.},
	title = "{Efficient generation of jets from magnetically arrested accretion on a rapidly spinning black hole}",
	journal = {\mnras},
	year = 2011,
	month = nov,
	volume = {418},
	number = {1},
	pages = {L79-L83},
	doi = {10.1111/j.1745-3933.2011.01147.x},
	archivePrefix = {arXiv},
	eprint = {1108.0412},
	primaryClass = {astro-ph.HE},
	adsurl = {https://ui.adsabs.harvard.edu/abs/2011MNRAS.418L..79T}
}

@ARTICLE{Porth2019,
       author = {{Porth}, Oliver and {Chatterjee}, Koushik and {Narayan}, Ramesh and {Gammie}, Charles F. and {Mizuno}, Yosuke and {Anninos}, Peter and {Baker}, John G. and {Bugli}, Matteo and {Chan}, Chi-kwan and {Davelaar}, Jordy and {Del Zanna}, Luca and {Etienne}, Zachariah B. and {Fragile}, P. Chris and {Kelly}, Bernard J. and {Liska}, Matthew and {Markoff}, Sera and {McKinney}, Jonathan C. and {Mishra}, Bhupendra and {Noble}, Scott C. and {Olivares}, H{\'e}ctor and {Prather}, Ben and {Rezzolla}, Luciano and {Ryan}, Benjamin R. and {Stone}, James M. and {Tomei}, Niccol{\`o} and {White}, Christopher J. and {Younsi}, Ziri and {Akiyama}, Kazunori and {Alberdi}, Antxon and {Alef}, Walter and {Asada}, Keiichi and {Azulay}, Rebecca and {Baczko}, Anne-Kathrin and {Ball}, David and {Balokovi{\'c}}, Mislav and {Barrett}, John and {Bintley}, Dan and {Blackburn}, Lindy and {Boland}, Wilfred and {Bouman}, Katherine L. and {Bower}, Geoffrey C. and {Bremer}, Michael and {Brinkerink}, Christiaan D. and {Brissenden}, Roger and {Britzen}, Silke and {Broderick}, Avery E. and {Broguiere}, Dominique and {Bronzwaer}, Thomas and {Byun}, Do-Young and {Carlstrom}, John E. and {Chael}, Andrew and {Chatterjee}, Shami and {Chen}, Ming-Tang and {Chen}, Yongjun and {Cho}, Ilje and {Christian}, Pierre and {Conway}, John E. and {Cordes}, James M. and {Geoffrey} and {Crew}, B. and {Cui}, Yuzhu and {De Laurentis}, Mariafelicia and {Deane}, Roger and {Dempsey}, Jessica and {Desvignes}, Gregory and {Doeleman}, Sheperd S. and {Eatough}, Ralph P. and {Falcke}, Heino and {Fish}, Vincent L. and {Fomalont}, Ed and {Fraga-Encinas}, Raquel and {Freeman}, Bill and {Friberg}, Per and {Fromm}, Christian M. and {G{\'o}mez}, Jos{\'e} L. and {Galison}, Peter and {Garc{\'\i}a}, Roberto and {Gentaz}, Olivier and {Georgiev}, Boris and {Goddi}, Ciriaco and {Gold}, Roman and {Gu}, Minfeng and {Gurwell}, Mark and {Hada}, Kazuhiro and {Hecht}, Michael H. and {Hesper}, Ronald and {Ho}, Luis C. and {Ho}, Paul and {Honma}, Mareki and {Huang}, Chih-Wei L. and {Huang}, Lei and {Hughes}, David H. and {Ikeda}, Shiro and {Inoue}, Makoto and {Issaoun}, Sara and {James}, David J. and {Jannuzi}, Buell T. and {Janssen}, Michael and {Jeter}, Britton and {Jiang}, Wu and {Johnson}, Michael D. and {Jorstad}, Svetlana and {Jung}, Taehyun and {Karami}, Mansour and {Karuppusamy}, Ramesh and {Kawashima}, Tomohisa and {Keating}, Garrett K. and {Kettenis}, Mark and {Kim}, Jae-Young and {Kim}, Junhan and {Kim}, Jongsoo and {Kino}, Motoki and {Koay}, Jun Yi and {Patrick} and {Koch}, M. and {Koyama}, Shoko and {Kramer}, Michael and {Kramer}, Carsten and {Krichbaum}, Thomas P. and {Kuo}, Cheng-Yu and {Lauer}, Tod R. and {Lee}, Sang-Sung and {Li}, Yan-Rong and {Li}, Zhiyuan and {Lindqvist}, Michael and {Liu}, Kuo and {Liuzzo}, Elisabetta and {Lo}, Wen-Ping and {Lobanov}, Andrei P. and {Loinard}, Laurent and {Lonsdale}, Colin and {Lu}, Ru-Sen and {MacDonald}, Nicholas R. and {Mao}, Jirong and {Marrone}, Daniel P. and {Marscher}, Alan P. and {Mart{\'\i}-Vidal}, Iv{\'a}n and {Matsushita}, Satoki and {Matthews}, Lynn D. and {Medeiros}, Lia and {Menten}, Karl M. and {Mizuno}, Izumi and {Moran}, James M. and {Moriyama}, Kotaro and {Moscibrodzka}, Monika and {M{\"u}ller}, Cornelia and {Nagai}, Hiroshi and {Nagar}, Neil M. and {Nakamura}, Masanori and {Narayanan}, Gopal and {Natarajan}, Iniyan and {Neri}, Roberto and {Ni}, Chunchong and {Noutsos}, Aristeidis and {Okino}, Hiroki and {Oyama}, Tomoaki and {{\"O}zel}, Feryal and {Palumbo}, Daniel C.~M. and {Patel}, Nimesh and {Pen}, Ue-Li and {Pesce}, Dominic W. and {Pi{\'e}tu}, Vincent and {Plambeck}, Richard and {PopStefanija}, Aleksandar and {Preciado-L{\'o}pez}, Jorge A. and {Psaltis}, Dimitrios and {Pu}, Hung-Yi and {Ramakrishnan}, Venkatessh and {Rao}, Ramprasad and {Rawlings}, Mark G. and {Raymond}, Alexander W. and {Ripperda}, Bart and {Roelofs}, Freek and {Rogers}, Alan and {Ros}, Eduardo and {Rose}, Mel and {Roshanineshat}, Arash and {Rottmann}, Helge and {Roy}, Alan L. and {Ruszczyk}, Chet and {Rygl}, Kazi L.~J. and {S{\'a}nchez}, Salvador and {S{\'a}nchez-Arguelles}, David and {Sasada}, Mahito and {Savolainen}, Tuomas and {Schloerb}, F. Peter and {Schuster}, Karl-Friedrich and {Shao}, Lijing and {Shen}, Zhiqiang and {Small}, Des and {Sohn}, Bong Won and {SooHoo}, Jason and {Tazaki}, Fumie and {Tiede}, Paul and {Tilanus}, Remo P.~J. and {Titus}, Michael and {Toma}, Kenji and {Torne}, Pablo and {Trent}, Tyler and {Trippe}, Sascha},
        title = "{The Event Horizon General Relativistic Magnetohydrodynamic Code Comparison Project}",
      journal = {\apjs},
         year = 2019,
        month = aug,
       volume = {243},
       number = {2},
          eid = {26},
        pages = {26},
          doi = {10.3847/1538-4365/ab29fd},
archivePrefix = {arXiv},
       eprint = {1904.04923},
 primaryClass = {astro-ph.HE},
       adsurl = {https://ui.adsabs.harvard.edu/abs/2019ApJS..243...26P}
}

@ARTICLE{Galishnikova2023,
       author = {{Galishnikova}, Alisa and {Philippov}, Alexander and {Quataert}, Eliot},
        title = "{Polarized Anisotropic Synchrotron Emission and Absorption and Its Application to Black Hole Imaging}",
      journal = {\apj},
         year = 2023,
        month = nov,
       volume = {957},
       number = {2},
          eid = {103},
        pages = {103},
          doi = {10.3847/1538-4357/acfa77},
archivePrefix = {arXiv},
       eprint = {2309.10029},
 primaryClass = {astro-ph.HE},
       adsurl = {https://ui.adsabs.harvard.edu/abs/2023ApJ...957..103G}
}

@ARTICLE{Neumayer_2010,
       author = {{Neumayer}, Nadine},
        title = "{The Supermassive Black Hole at the Heart of Centaurus A: Revealed by the Kinematics of Gas and Stars}",
      journal = {\pasa},
         year = 2010,
        month = oct,
       volume = {27},
       number = {4},
        pages = {449-456},
          doi = {10.1071/AS09080},
archivePrefix = {arXiv},
       eprint = {1002.0965},
 primaryClass = {astro-ph.CO},
       adsurl = {https://ui.adsabs.harvard.edu/abs/2010PASA...27..449N}
}

@ARTICLE{Harris_2010,
       author = {{Harris}, Gretchen L.~H. and {Rejkuba}, Marina and {Harris}, William E.},
        title = "{The Distance to NGC 5128 (Centaurus A)}",
      journal = {\pasa},
         year = 2010,
        month = oct,
       volume = {27},
       number = {4},
        pages = {457-462},
          doi = {10.1071/AS09061},
archivePrefix = {arXiv},
       eprint = {0911.3180},
 primaryClass = {astro-ph.GA},
       adsurl = {https://ui.adsabs.harvard.edu/abs/2010PASA...27..457H}
}

@ARTICLE{Kawazura_2020,
       author = {{Kawazura}, Y. and {Schekochihin}, A.~A. and {Barnes}, M. and {TenBarge}, J.~M. and {Tong}, Y. and {Klein}, K.~G. and {Dorland}, W.},
        title = "{Ion versus Electron Heating in Compressively Driven Astrophysical Gyrokinetic Turbulence}",
      journal = {Physical Review X},
         year = 2020,
        month = oct,
       volume = {10},
       number = {4},
          eid = {041050},
        pages = {041050},
          doi = {10.1103/PhysRevX.10.041050},
archivePrefix = {arXiv},
       eprint = {2004.04922},
 primaryClass = {physics.plasm-ph},
       adsurl = {https://ui.adsabs.harvard.edu/abs/2020PhRvX..10d1050K}
}

@ARTICLE{Pandya_2016,
       author = {{Pandya}, Alex and {Zhang}, Zhaowei and {Chandra}, Mani and {Gammie}, Charles F.},
        title = "{Polarized Synchrotron Emissivities and Absorptivities for Relativistic Thermal, Power-law, and Kappa Distribution Functions}",
      journal = {\apj},
         year = 2016,
        month = may,
       volume = {822},
       number = {1},
          eid = {34},
        pages = {34},
          doi = {10.3847/0004-637X/822/1/34},
archivePrefix = {arXiv},
       eprint = {1602.08749},
 primaryClass = {astro-ph.HE},
       adsurl = {https://ui.adsabs.harvard.edu/abs/2016ApJ...822...34P}
}

@ARTICLE{Meringolo_2023,
       author = {{Meringolo}, Claudio and {Cruz-Osorio}, Alejandro and {Rezzolla}, Luciano and {Servidio}, Sergio},
        title = "{Microphysical Plasma Relations from Special-relativistic Turbulence}",
      journal = {\apj},
         year = 2023,
        month = feb,
       volume = {944},
       number = {2},
          eid = {122},
        pages = {122},
          doi = {10.3847/1538-4357/acaefe},
archivePrefix = {arXiv},
       eprint = {2301.02669},
 primaryClass = {astro-ph.HE},
       adsurl = {https://ui.adsabs.harvard.edu/abs/2023ApJ...944..122M}
}

@article{Zhdankin_2023,
   title={Synchrotron Firehose Instability},
   volume={944},
   ISSN={1538-4357},
   url={http://dx.doi.org/10.3847/1538-4357/acaf54},
   DOI={10.3847/1538-4357/acaf54},
   number={1},
   journal={\apj},
   publisher={American Astronomical Society},
   author={Zhdankin, Vladimir and Kunz, Matthew W. and Uzdensky, Dmitri A.},
   year={2023},
   month=feb, pages={24} }

@ARTICLE{Mizuno2021,
       author = {{Mizuno}, Yosuke and {Fromm}, Christian M. and {Younsi}, Ziri and {Porth}, Oliver and {Olivares}, Hector and {Rezzolla}, Luciano},
        title = "{Comparison of the ion-to-electron temperature ratio prescription: GRMHD simulations with electron thermodynamics}",
      journal = {\mnras},
         year = 2021,
        month = sep,
       volume = {506},
       number = {1},
        pages = {741-758},
          doi = {10.1093/mnras/stab1753},
archivePrefix = {arXiv},
       eprint = {2106.09272},
 primaryClass = {astro-ph.HE},
       adsurl = {https://ui.adsabs.harvard.edu/abs/2021MNRAS.506..741M}
}

@article{Tsunetoe_2025,
doi = {10.3847/1538-4357/adc37a},
url = {https://dx.doi.org/10.3847/1538-4357/adc37a},
year = {2025},
month = {apr},
publisher = {The American Astronomical Society},
volume = {984},
number = {1},
pages = {35},
author = {Tsunetoe, Yuh and Pesce, Dominic W. and Narayan, Ramesh and Chael, Andrew and Gelles, Zachary and Gammie, Charles and Quataert, Eliot and Palumbo, Daniel},
title = {Limb-brightened Jet in M87 from Anisotropic Nonthermal Electrons},
journal = {\apj}
}

@ARTICLE{Melrose_1971,
       author = {{Melrose}, D.~B.},
        title = "{On the Degree of Circular Polarization of Synchrotron Radiation}",
      journal = {\apss},
         year = 1971,
        month = jul,
       volume = {12},
       number = {1},
        pages = {172-192},
          doi = {10.1007/BF00656148},
       adsurl = {https://ui.adsabs.harvard.edu/abs/1971Ap&SS..12..172M}
}

@ARTICLE{Goddi2021,
       author = {{Goddi}, Ciriaco and {Mart{\'\i}-Vidal}, Iv{\'a}n and {Messias}, Hugo and {Bower}, Geoffrey C. and {Broderick}, Avery E. and {Dexter}, Jason and {Marrone}, Daniel P. and {Moscibrodzka}, Monika and {Nagai}, Hiroshi and {Algaba}, Juan Carlos and {Asada}, Keiichi and {Crew}, Geoffrey B. and {G{\'o}mez}, Jos{\'e} L. and {Impellizzeri}, C.~M. Violette and {Janssen}, Michael and {Kadler}, Matthias and {Krichbaum}, Thomas P. and {Lico}, Rocco and {Matthews}, Lynn D. and {Nathanail}, Antonios and {Ricarte}, Angelo and {Ros}, Eduardo and {Younsi}, Ziri and {Akiyama}, Kazunori and {Alberdi}, Antxon and {Alef}, Walter and {Anantua}, Richard and {Azulay}, Rebecca and {Baczko}, Anne-Kathrin and {Ball}, David and {Balokovi{\'c}}, Mislav and {Barrett}, John and {Benson}, Bradford A. and {Bintley}, Dan and {Blackburn}, Lindy and {Blundell}, Raymond and {Boland}, Wilfred and {Bouman}, Katherine L. and {Boyce}, Hope and {Bremer}, Michael and {Brinkerink}, Christiaan D. and {Brissenden}, Roger and {Britzen}, Silke and {Broguiere}, Dominique and {Bronzwaer}, Thomas and {Byun}, Do-Young and {Carlstrom}, John E. and {Chael}, Andrew and {Chan}, Chi-kwan and {Chatterjee}, Shami and {Chatterjee}, Koushik and {Chen}, Ming-Tang and {Chen}, Yongjun and {Chesler}, Paul M. and {Cho}, Ilje and {Christian}, Pierre and {Conway}, John E. and {Cordes}, James M. and {Crawford}, Thomas M. and {Cruz-Osorio}, Alejandro and {Cui}, Yuzhu and {Davelaar}, Jordy and {De Laurentis}, Mariafelicia and {Deane}, Roger and {Dempsey}, Jessica and {Desvignes}, Gregory and {Doeleman}, Sheperd S. and {Eatough}, Ralph P. and {Falcke}, Heino and {Farah}, Joseph and {Fish}, Vincent L. and {Fomalont}, Ed and {Ford}, H. Alyson and {Fraga-Encinas}, Raquel and {Freeman}, William T. and {Friberg}, Per and {Fromm}, Christian M. and {Fuentes}, Antonio and {Galison}, Peter and {Gammie}, Charles F. and {Garc{\'\i}a}, Roberto and {Gentaz}, Olivier and {Georgiev}, Boris and {Gold}, Roman and {G{\'o}mez-Ruiz}, Arturo I. and {Gu}, Minfeng and {Gurwell}, Mark and {Hada}, Kazuhiro and {Haggard}, Daryl and {Hecht}, Michael H. and {Hesper}, Ronald and {Ho}, Luis C. and {Ho}, Paul and {Honma}, Mareki and {Huang}, Chih-Wei L. and {Huang}, Lei and {Hughes}, David H. and {Inoue}, Makoto and {Issaoun}, Sara and {James}, David J. and {Jannuzi}, Buell T. and {Jeter}, Britton and {Jiang}, Wu and {Jimenez-Rosales}, Alejandra and {Johnson}, Michael D. and {Jorstad}, Svetlana and {Jung}, Taehyun and {Karami}, Mansour and {Karuppusamy}, Ramesh and {Kawashima}, Tomohisa and {Keating}, Garrett K. and {Kettenis}, Mark and {Kim}, Dong-Jin and {Kim}, Jae-Young and {Kim}, Jongsoo and {Kim}, Junhan and {Kino}, Motoki and {Koay}, Jun Yi and {Kofuji}, Yutaro and {Koch}, Patrick M. and {Koyama}, Shoko and {Kramer}, Michael and {Kramer}, Carsten and {Kuo}, Cheng-Yu and {Lauer}, Tod R. and {Lee}, Sang-Sung and {Levis}, Aviad and {Li}, Yan-Rong and {Li}, Zhiyuan and {Lindqvist}, Michael and {Lindahl}, Greg and {Liu}, Jun and {Liu}, Kuo and {Liuzzo}, Elisabetta and {Lo}, Wen-Ping and {Lobanov}, Andrei P. and {Loinard}, Laurent and {Lonsdale}, Colin and {Lu}, Ru-Sen and {MacDonald}, Nicholas R. and {Mao}, Jirong and {Marchili}, Nicola and {Markoff}, Sera and {Marscher}, Alan P. and {Matsushita}, Satoki and {Medeiros}, Lia and {Menten}, Karl M. and {Mizuno}, Izumi and {Mizuno}, Yosuke and {Moran}, James M. and {Moriyama}, Kotaro and {M{\"u}ller}, Cornelia and {Musoke}, Gibwa and {Mej{\'\i}as}, Alejandro Mus and {Nagar}, Neil M. and {Nakamura}, Masanori and {Narayan}, Ramesh and {Narayanan}, Gopal and {Natarajan}, Iniyan and {Neilsen}, Joey and {Neri}, Roberto and {Ni}, Chunchong and {Noutsos}, Aristeidis and {Nowak}, Michael A. and {Okino}, Hiroki and {Olivares}, H{\'e}ctor and {Ortiz-Le{\'o}n}, Gisela N. and {Oyama}, Tomoaki and {{\"O}zel}, Feryal and {Palumbo}, Daniel C.~M. and {Park}, Jongho and {Patel}, Nimesh and {Pen}, Ue-Li and {Pesce}, Dominic W. and {Pi{\'e}tu}, Vincent and {Plambeck}, Richard and {PopStefanija}, Aleksandar and {Porth}, Oliver and {P{\"o}tzl}, Felix M. and {Prather}, Ben and {Preciado-L{\'o}pez}, Jorge A. and {Psaltis}, Dimitrios and {Pu}, Hung-Yi and {Ramakrishnan}, Venkatessh and {Rao}, Ramprasad and {Rawlings}, Mark G. and {Raymond}, Alexander W. and {Rezzolla}, Luciano and {Ripperda}, Bart and {Roelofs}, Freek and {Rogers}, Alan and {Rose}, Mel and {Roshanineshat}, Arash and {Rottmann}, Helge and {Roy}, Alan L. and {Ruszczyk}, Chet and {Rygl}, Kazi L.~J. and {S{\'a}nchez}, Salvador and {S{\'a}nchez-Arguelles}, David and {Sasada}, Mahito},
        title = "{Polarimetric Properties of Event Horizon Telescope Targets from ALMA}",
      journal = {\apjl},
         year = 2021,
        month = mar,
       volume = {910},
       number = {1},
          eid = {L14},
        pages = {L14},
          doi = {10.3847/2041-8213/abee6a},
archivePrefix = {arXiv},
       eprint = {2105.02272},
 primaryClass = {astro-ph.GA},
       adsurl = {https://ui.adsabs.harvard.edu/abs/2021ApJ...910L..14G}
}

@ARTICLE{Blandford1979,
       author = {{Blandford}, R.~D. and {K{\"o}nigl}, A.},
        title = "{Relativistic jets as compact radio sources.}",
      journal = {\apj},
         year = 1979,
        month = aug,
       volume = {232},
        pages = {34-48},
          doi = {10.1086/157262},
       adsurl = {https://ui.adsabs.harvard.edu/abs/1979ApJ...232...34B}
}

@ARTICLE{Mahdi2024,
       author = {{Najafi-Ziyazi}, Mahdi and {Davelaar}, Jordy and {Mizuno}, Yosuke and {Porth}, Oliver},
        title = "{Flares in the Galactic centre - II. Polarization signatures of flares at mm-wavelengths}",
      journal = {\mnras},
         year = 2024,
        month = jul,
       volume = {531},
       number = {4},
        pages = {3961-3972},
          doi = {10.1093/mnras/stae1343},
archivePrefix = {arXiv},
       eprint = {2308.16740},
 primaryClass = {astro-ph.HE},
       adsurl = {https://ui.adsabs.harvard.edu/abs/2024MNRAS.531.3961N}
}

@ARTICLE{Porth2021,
       author = {{Porth}, O. and {Mizuno}, Y. and {Younsi}, Z. and {Fromm}, C.~M.},
        title = "{Flares in the Galactic Centre - I. Orbiting flux tubes in magnetically arrested black hole accretion discs}",
      journal = {\mnras},
         year = 2021,
        month = apr,
       volume = {502},
       number = {2},
        pages = {2023-2032},
          doi = {10.1093/mnras/stab163},
archivePrefix = {arXiv},
       eprint = {2006.03658},
 primaryClass = {astro-ph.HE},
       adsurl = {https://ui.adsabs.harvard.edu/abs/2021MNRAS.502.2023P}
}

@ARTICLE{Sinnis2023,
       author = {{Sinnis}, Charalampos and {Vlahakis}, Nektarios},
        title = "{Linear stability analysis of relativistic magnetized jets. The Kelvin-Helmholtz mode}",
      journal = {\aap},
         year = 2023,
        month = dec,
       volume = {680},
          eid = {A46},
        pages = {A46},
          doi = {10.1051/0004-6361/202347647},
       adsurl = {https://ui.adsabs.harvard.edu/abs/2023A&A...680A..46S}
}

@ARTICLE{Hardee2007,
       author = {{Hardee}, Philip E.},
        title = "{Stability Properties of Strongly Magnetized Spine-Sheath Relativistic Jets}",
      journal = {\apj},
         year = 2007,
        month = jul,
       volume = {664},
       number = {1},
        pages = {26-46},
          doi = {10.1086/518409},
archivePrefix = {arXiv},
       eprint = {0704.1621},
 primaryClass = {astro-ph},
       adsurl = {https://ui.adsabs.harvard.edu/abs/2007ApJ...664...26H}
}

@ARTICLE{Lyutikov2005,
       author = {{Lyutikov}, M. and {Pariev}, V.~I. and {Gabuzda}, D.~C.},
        title = "{Polarization and structure of relativistic parsec-scale AGN jets}",
      journal = {\mnras},
         year = 2005,
        month = jul,
       volume = {360},
       number = {3},
        pages = {869-891},
          doi = {10.1111/j.1365-2966.2005.08954.x},
archivePrefix = {arXiv},
       eprint = {astro-ph/0406144},
 primaryClass = {astro-ph},
       adsurl = {https://ui.adsabs.harvard.edu/abs/2005MNRAS.360..869L}
}

@article{Tingay_1998,
doi = {10.1086/300257},
url = {https://doi.org/10.1086/300257},
year = {1998},
month = {mar},
publisher = {},
volume = {115},
number = {3},
pages = {960},
author = {Tingay, S. J. and Jauncey, D. L. and Reynolds, J. E. and Tzioumis, A. K. and King, E. A. and Preston, R. A. and Jones, D. L. and Murphy, D. W. and Meier, D. L. and van Ommen, T. D. and McCulloch, P. M. and Ellingsen, S. P. and Costa, M. E. and Edwards, P. G. and Lovell, J. E. J. and Nicolson, G. D. and Quick, J. F. H. and Kemball, A. J. and Migenes, V. and Harbison, P. and Jones, P. A. and White, G. L. and Gough, R. G. and Ferris, R. H. and Sinclair, M. W. and Clay, R. W.},
title = {The Subparsec-Scale Structure and Evolution of Centaurus A:
The Nearest Active Radio Galaxy},
journal = {The Astronomical Journal}
}

@ARTICLE{Glaser2026,
       author = {{Glaser}, Felix and {Fromm}, Christian M. and {Mizuno}, Yosuke and {Kadler}, Matthias and {Mannheim}, Karl},
        title = "{The Magnetic Filling in MAD Simulations and its Impact on the Jet in M87}",
      journal = {arXiv e-prints},
         year = 2026,
        month = feb,
          eid = {arXiv:2602.20893},
        pages = {arXiv:2602.20893},
          doi = {10.48550/arXiv.2602.20893},
archivePrefix = {arXiv},
       eprint = {2602.20893},
 primaryClass = {astro-ph.HE},
       adsurl = {https://ui.adsabs.harvard.edu/abs/2026arXiv260220893G}
}

\begin{acknowledgements}
This research is supported by the DFG research grant ``Jet physics on horizon scales and beyond" (Grant No.  443220636) within the DFG research unit ``Relativistic Jets in Active Galaxies" (FOR 5195). The numerical simulations and calculations have been performed on \texttt{MISTRAL} at the Chair of Astronomy at the JMU Wuerzburg. The numerical resources for F.G. have been provided by a study grant from the Faculty of Physics at JMU Wuerzburg.
Y.M. acknowledges support from the National Key R\&D Program of China (grant No. 2023YFE0101200), the National Natural Science Foundation of China (grant No. 12273022, 12511540053), and the Shanghai Municipality Orientation Program of Basic Research for International Scientists (grant No. 22JC1410600).
\end{acknowledgements}

\begin{appendix}

\section{Pitch-angle dependence of anisotropic emissivities and absorptivities}

\begin{figure}[h!]
\centering
\includegraphics[width=\linewidth]{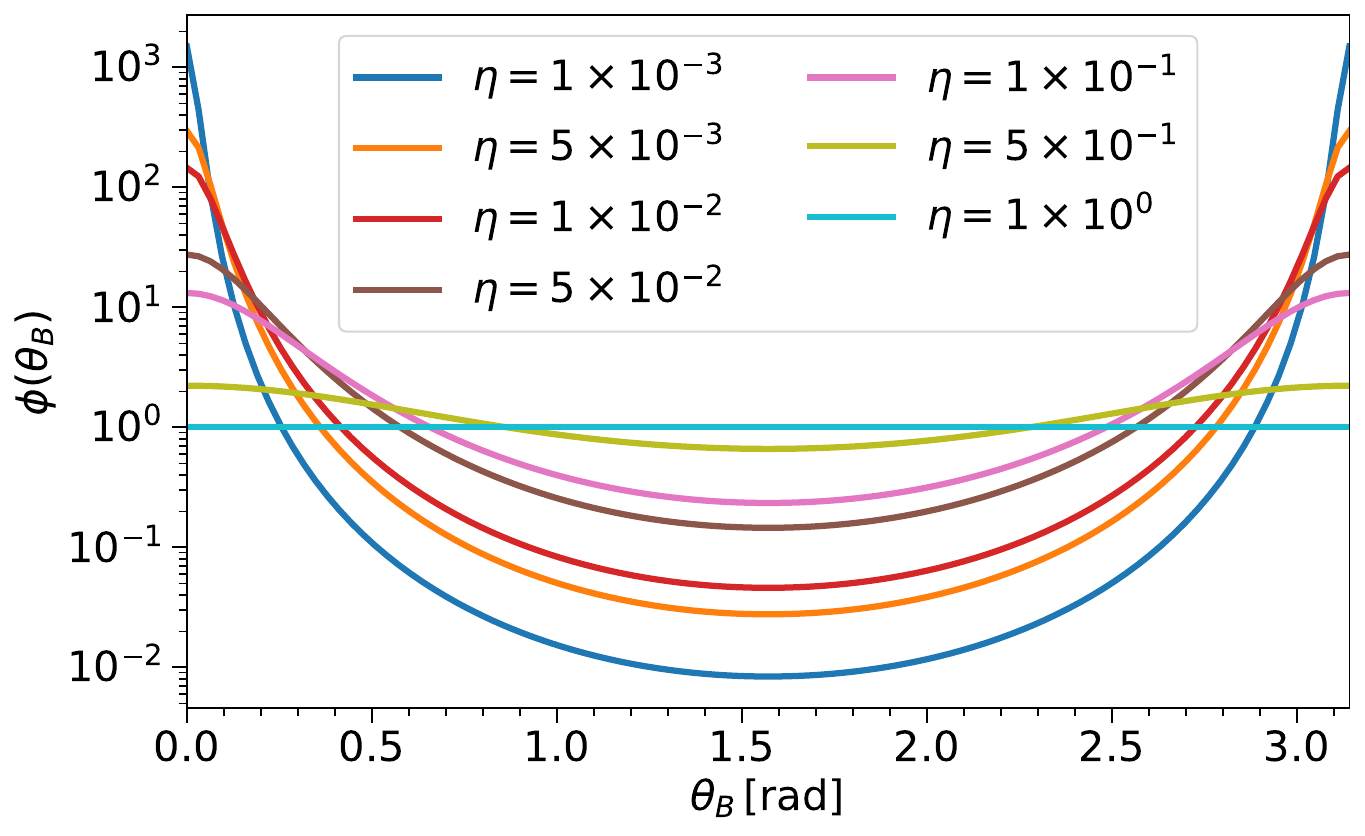}
\caption{Illustration of the pitch angle anisotropy for different values of $\eta$}
\label{fig:aniso}
\end{figure}

In Fig.~\ref{fig:aniso}, we illustrate the effect of pitch-angle  $\theta_{\rm B}$ on the degree of anisotropy of the emissivity/absorptivity distributions in $\{\mathcal{I},\mathcal{Q},\mathcal{U}\}$ for different values of $\eta$, while keeping $p=3.5$ ($\kappa=2.5$) fixed (see Eq. \ref{eqn:anisoja}). We find that decreasing $\eta$ leads to enhanced values of $\theta_{\rm B}$ at the edges of the distribution, i.e., at small and large pitch angles, when compared to the isotropic case ($\eta=1$, cyan line).
    
\end{appendix}

\end{document}